\documentclass[preprint]{revtex4-2}

\usepackage{fancyhdr}
\usepackage{amsmath,amssymb,amsfonts}
\usepackage{graphicx}
\usepackage{citesort}
\usepackage{xcolor}
\usepackage{comment}
\usepackage[normalem]{ulem}
\usepackage[colorlinks=true, linkcolor=blue, citecolor=red, urlcolor=green]{hyperref}
\usepackage{lineno}

\begin{document}

\title{Unconventional Growth Kinetics and Fractal Interfaces of Colloidal Phase Separation in Active Liquids}

\author{Pragya Kushwaha$^{1}$}
\author{Pratikshya Jena${^2}$} 
\author{Partha Sarathi Mondal${^2}$} 
\author{Nayanthara J R $^{1}$}
\author{Sanjay Puri$^3$} 
\author{Shradha Mishra${^2}$} 
\author{Vijayakumar Chikkadi$^{1}$}
\thanks{Corresponding authors : purijnu@gmail.com, smishra.phy@itbhu.ac.in, vijayck@iiserpune.ac.in}

\affiliation{$^{1}$ Indian Institute of Science Education and Research Pune, India 411008}
\affiliation{$^{2}$ Indian Institute of Technology (BHU) Varanasi, India 221005}
\affiliation{$^{3}$ School of Physical Sciences, Jawaharlal Nehru University, New Delhi, India 110067}

\begin{abstract}

Phase separation driven by nonequilibrium fluctuations is a hallmark of both living and synthetic active matter. Unlike equilibrium systems,  where ordered states arise from the minimization of free energy, active systems are fueled by a constant injection of energy at the microscopic scale. The emergence of ordered phases in such driven systems challenges our conventional views of domain growth and interfacial structure. In this study, we investigate the coarsening of colloidal clusters in active liquids containing E. coli. Our experiments reveal that uniform dispersions of colloids and swimmers are inherently unstable, resulting in spontaneous phase separation characterized by fractal interfaces and unconventional kinetics. The correlation function of the order parameter displays dynamical scaling, with the size of colloidal domains initially growing as $t^{1/z}$, where $z \sim 4$, in contrast to the well-known growth laws for thermal systems with a conserved order parameter. Furthermore, the structure factor exhibits non-Porod behavior, indicating domains with fractal interfaces. This non-Porod behavior also manifests \sout{itself} as a cusp singularity in the correlation function. We elucidate our experimental findings using a scalar field theory in which the nonequilibrium fluctuations arising from swimmer activity are modeled as spatio-temporally correlated noise. It quantitatively reproduces the domain growth law and non-Porod structure factor resulting from fractal interfaces observed in experiments. In addition, it also reveals a fluctuating microphase separation, where the initial growth of the domain is eventually arrested, thus shedding new light on the microscopic origins of the unconventional phase separation of colloids in active liquids.
\end{abstract}

\maketitle

\section{Introduction}

Homogeneous binary mixtures that are thermodynamically stable at high temperatures become unstable when cooled below the coexistence curve. This instability leads to the formation of domains of ordered phases, which grow over time and eventually result in a stable, phase-separated state \cite{Cahn58, Cahn61}. This fascinating phenomenon is prevalent in nature and has relevance in various fields of physics, materials science, biology, and everyday life \cite{Bray94, Puri09}. In recent years, there has been increasing interest in nonequilibrium aspects of phase separation in both living and synthetic active systems \cite{Nardini24}. Examples include living matter such as colonies of bacteria \cite{Shaevitz19,Poon12,Tailleur10,Huang20}, biological tissues \cite{Steinberg63,Steinberg63,Trivedi22}, and biomolecular condensates within cells \cite{Julicher14,Rosen17,Lee19,Haataja18,Pappu15}, as well as synthetic active matter such as self-propelled colloidal particles \cite{Dauchot19, Bocquet12, Chaikin13, Speck13, Stone18, Vermant20, Cottin18, Granick21,Luijten17} and mixtures of microtubules and motors \cite{Dogic22,Dogic22-1,Dogic22,Dogic23}. The central goal of many of these investigations is to understand the effect of activity on the nature of phase separation. In particular, how the activity influences the kinetics of phase separation, such as dynamic scaling, domain growth laws, and the nature of interfaces, remain unclear. \\

The answers to these questions have inspired novel experiments and coarse-grained models of active phase separation that include the effect of the persistent motion of active units. These aspects have been investigated theoretically \cite{Bialke-2013} within the framework of motility-induced phase separation (MIPS) \cite{Cates15-1,Cates-2008} where a density-dependent motility of active units results in a positive feedback mechanism between clustering and slowing of particles, leading to phase separation. In other words, particles slow down on encountering dense regions, leading to their accumulation, which in turn enhances slowing. This model was motivated by the manifestation of phase separation in a purely repulsive particle system of active Brownian particles (ABP) \cite{Baskaran13, Marchetti12} and experiments with active Janus colloids \cite{Chaikin13, Speck13, Bocquet12}. Recent efforts have focused on developing scalar field theories with a conserved order parameter \cite{Nardini24,Cate14,Yeomans24}, along the lines of the Cahn-Hilliard equation and Model B \cite{Bray94,Puri09}. These are active versions of equilibrium models, which are known as Active Model B \cite{Cate14,Halperin77}, Active Model B$+$ \cite{Cates22, Cates18} and others \cite{Yeomans24,Cates15-2}. The presence of activity breaks the principle of detailed balance and time-reversal symmetry. Such studies have uncovered unique features of active phase separation that are absent in equilibrium systems. These include deviations from the conventional Lifshitz-Slyozov (LS) domain growth law \cite{Cate14}, negative interfacial tension \cite{Speck15} leading to the reversal of the Ostwald process \cite{Cates18}, microphase separation \cite{Cates18-2} and other novel phenomena. The more recent proposal has been the noise-induced phase separation model \cite{Pagonabarraga24}, which uses Model B with temporally correlated noise to study the effect of an active bath on phase separation of passive species. There have also been several studies of active phase separation based on dynamical density functional theory (DDFT), including colloidal particles with arbitrary shape \cite{Lowen11, Lowen21}, “dry” and “wet” active matter \cite{Lowen16, Lowen21, Wittkowski23}, and more recent predictive local field theories for interacting active Brownian particles \cite{Wittkowski20, Wittkowski20-1}. In these approaches, the equations of motion are derived from the microscopic Langevin dynamics in the low-density limit. An important advantage of some of these methods is that they retain detailed information about particle interactions, allowing one to examine the effects of particle shape, interaction potentials, and other microscopic features. However, in most phase-separation problems, such microscopic details become irrelevant in the asymptotic long-time limit. \\

Besides these developments, several studies have investigated phase separation in active matter \cite{Dogic22, Baskaran13, Dogic22-1, Dogic23, Yeomans23, Yeomans24, Yeomans24-1, Cates19}, contrasting it with equilibrium scenarios. Simulations of active Brownian particles (ABPs) by Redner et al. \cite{Baskaran13} reported an unconventional cluster growth in the coarsening regime, characterized by a $t^{1/4}$ growth law. Similarly, simulations of Active Model B revealed a crossover in domain growth, from $t^{1/3}$ at early times to $t^{1/4}$ at later times, with the crossover timescale determined by the activity level \cite{Puri21}. In contrast to these findings, experimental studies with self-propelled Janus colloids \cite{Douchot19, Granick21} reported equilibrium-like growth with $t^{1/3}$ scaling. Likewise, the coarsening of colloidal clusters in active liquids displayed comparable behavior \cite{Auradou23}. Surprisingly, most experiments so far have reported growth kinetics that closely resemble those of conventional thermal systems. A notable exception comes from experiments on sticky colloidal spheres suspended in a chiral active bath of E. coli \cite{Palacci23}, which revealed a novel model of aggregation driving colloidal clusters into unconventional gel states that are unattainable under thermal equilibrium.\\

In a separate set of studies, the kinetics of coarsening and the structure of the interfaces were investigated in a system of particles sliding on a fluctuating surface using lattice models \cite{Barma00, Barma01, Barma23, Majumdar01}, as well as in simulations of the Vicsek model \cite{Puri20}, cooling granular gasses \cite{Rajesh07}, and the organization of passive molecules driven by molecular motors on cell surfaces \cite{Rao16}. These driven systems have revealed fluctuation-dominated phase ordering (FDPO), characterized by unusual coarsening kinetics and interfaces with non-Porod features. Observations of such features of phase ordering have so far been limited to numerical studies, and experimental evidence is absent. \\

Here, we show that colloidal assemblies driven in active liquids of E. coli display several novel features of nonequilibrium coarsening previously reported in numerical studies \cite{Barma00, Barma01, Barma23, Cate14, Puri21}, such as domain growth showing strong deviations from conventional growth laws for thermal systems with a conserved order parameter. We observe that $L(t) \sim t^{1/4}$ over a range of colloidal densities and bacterial concentrations. The structure factor reveals non-Porod signatures at large wave-vectors $k$, which also manifests itself as a cusp singularity at short distances in the corresponding two-point correlation function of the order parameter \cite{Barma00,Barma23,Puri14}. These features point to rough and diffuse interfaces, unlike the sharp interfaces found in the equilibrium counterparts. We further establish the fractal nature of the interfaces by directly calculating the fractal dimension of the interface from the real-space particle positions. We elucidate all these experimental observations using a scalar field theory where the nonequilibrium fluctuations arising from swimmer activity are modeled as spatio-temporally correlated noise. Such a coarse-grained model, which breaks time-reversal symmetry and detailed balance, successfully captures several features observed in our experiments. In addition our model predicts microphase separation of colloids driven by active fluctuation arising from swimmer motion, thus pointing to a novel fluctuation dominated ordering of colloids in active liquids. \\

\begin{figure}
    \centering
    \begin{tabular}{ll}
    \includegraphics[width=0.32\textwidth]{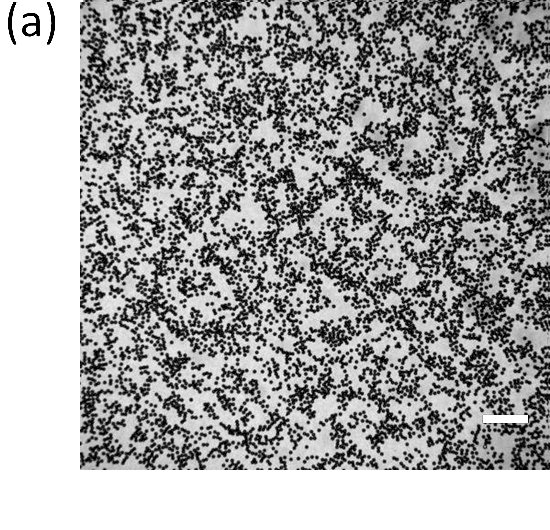} & 
    \includegraphics[width=0.32\textwidth]{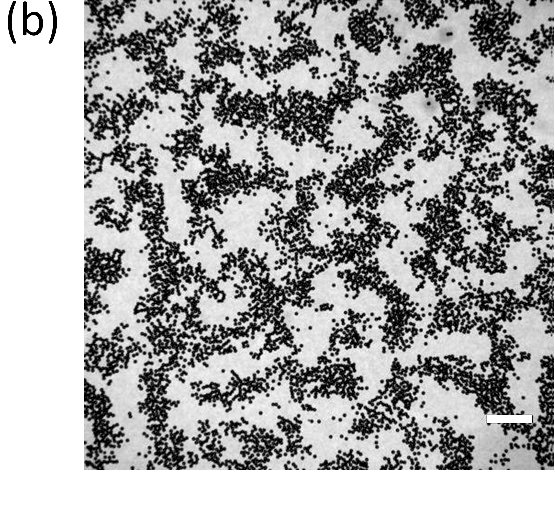}\\
    \includegraphics[width=0.32\textwidth]{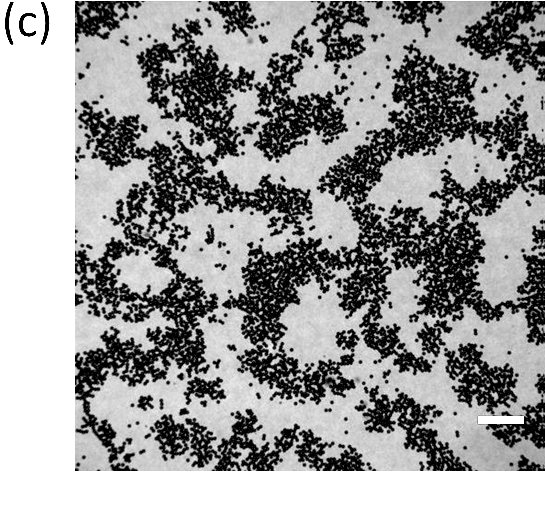} & 
    \includegraphics[width=0.37\textwidth]{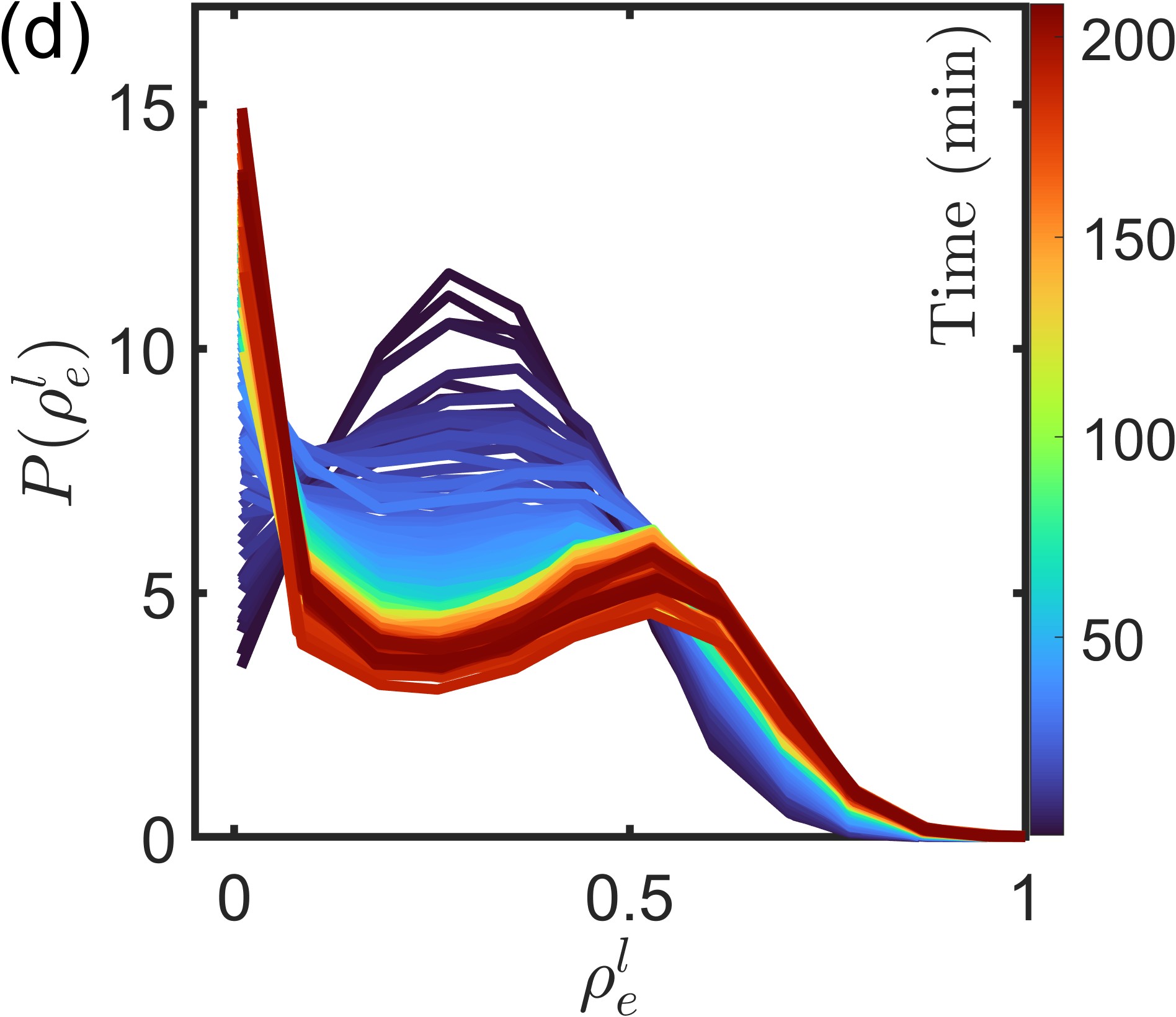}\\
    \end{tabular}
    \caption{Coarsening of colloidal clusters in active liquids. The panels (a)-(c) are bright field images of the suspension at different time intervals: 1 min (\textbf{a}), 40 min (\textbf{b}), and 210 min (\textbf{c}), respectively. The scale bar in the images is 200$\mu m$. (\textbf{d}) The distribution $P(\rho_e^l)$ of local density $\rho_e^l$. The time evolution of the distribution in minutes is color coded. The concentration of bacteria is $c_e=5b_0$ and the density of colloids is $\rho_e^0=0.28$.}
    \label{Fig1}
\end{figure}

\section{Coarsening kinetics of colloidal clusters}

Our experiments used suspensions containing a mixture of motile bacteria and colloidal beads to investigate the active phase separation of colloidal domains. We used polystyrene beads of diameter $\sigma = 15\mu m$ and E. coli cells \cite{Chikkadi23} as motile swimmers. The procedure followed for preparing the suspensions is described in the Methods section. The concentration of bacteria in most of the measurements is $c_e=5~b_0$, where $b_0 = 6 \times 10^9$ cells/ml. The dynamics of colloidal beads in our suspensions is driven by the inherent activity of swimmers. The colloidal beads and E. coli swimmers are mixed uniformly before being introduced into a sample chamber. This chamber is designed in a way to optimize the duration of our experiments; see Methods and section I-A and I-B of the Supplementary Information for details. Once the suspension is loaded into the sample chamber, the colloidal particles settle to the bottom of the chamber, and subsequently bright-field images are acquired at a speed of 2 FPS over a total duration of 210 minutes. All the physical quantities measured in our experiments are denoted with a subscript $e$. \\
 
We started our experiments in a homogeneous phase with colloidal particles dispersed uniformly in suspensions of microswimmers. The homogeneous phase of our mixture is unstable; therefore, it undergoes phase separation by forming dynamic clusters that grow over time. Recent investigations have shown that the interplay of effective interactions between colloids and swimmer activity plays a key role in the formation of dynamic clusters \cite{Chikkadi23, Fakhri22}. The interaction strength can be tuned by adjusting the size ratio between colloidal particles and swimmers \cite{Chikkadi23}. In our current experiments, the size ratio of the colloidal beads and swimmers is set to $S\sim 5.5$. The phase ordering that follows is depicted using bright-field images of colloidal particles in Figs.~\ref{Fig1}(a)-(c) at different time intervals corresponding to $t = 1$ min, $t = 40$ min and $t = 210$ min, respectively. The average density of colloids is $\rho_e^0=0.28$ and the size of the scale bar in the images is 200 $\mu$m. A video of the coarsening process is presented in Supplementary Video SM1. Clearly, the colloids form clusters that increase in size as time progresses. To quantify the evolving structures, we computed the distribution of the coarse-grained local density $\rho_e^l(\mathbf{r},t)$ of the particles using a box size of $2\sigma \times 2\sigma$. Colloidal particles were tracked using standard methods \cite{Grier96}, see Methods for details. The resulting distribution $P(\rho_e^l)$ obtained from the particle positions is shown in Fig.~\ref{Fig1}(d). The distribution is unimodal in the early stages of coarsening; however, it becomes bimodal in later stages with two peaks at $\rho_e^l\sim0$ and $\rho_e^l\sim0.6$, corresponding to the colloid-poor and colloid-rich phases, respectively. The color indicates the progression of time in minutes. Although our experimental time window allows us to investigate several aspects of the phase-separation dynamics, it is not long enough to access the steady-state behavior at long times. Furthermore, colloids of smaller size ratios lead to smaller effective interaction, so they do not exhibit strong dynamic clustering, see Supplementary Video SM2 for colloids of size ratio $S\sim 3$.\\

\begin{figure}[t]
    \centering
    \begin{tabular}{ccc}
        \includegraphics[width=0.29\textwidth,height=0.25\textwidth]{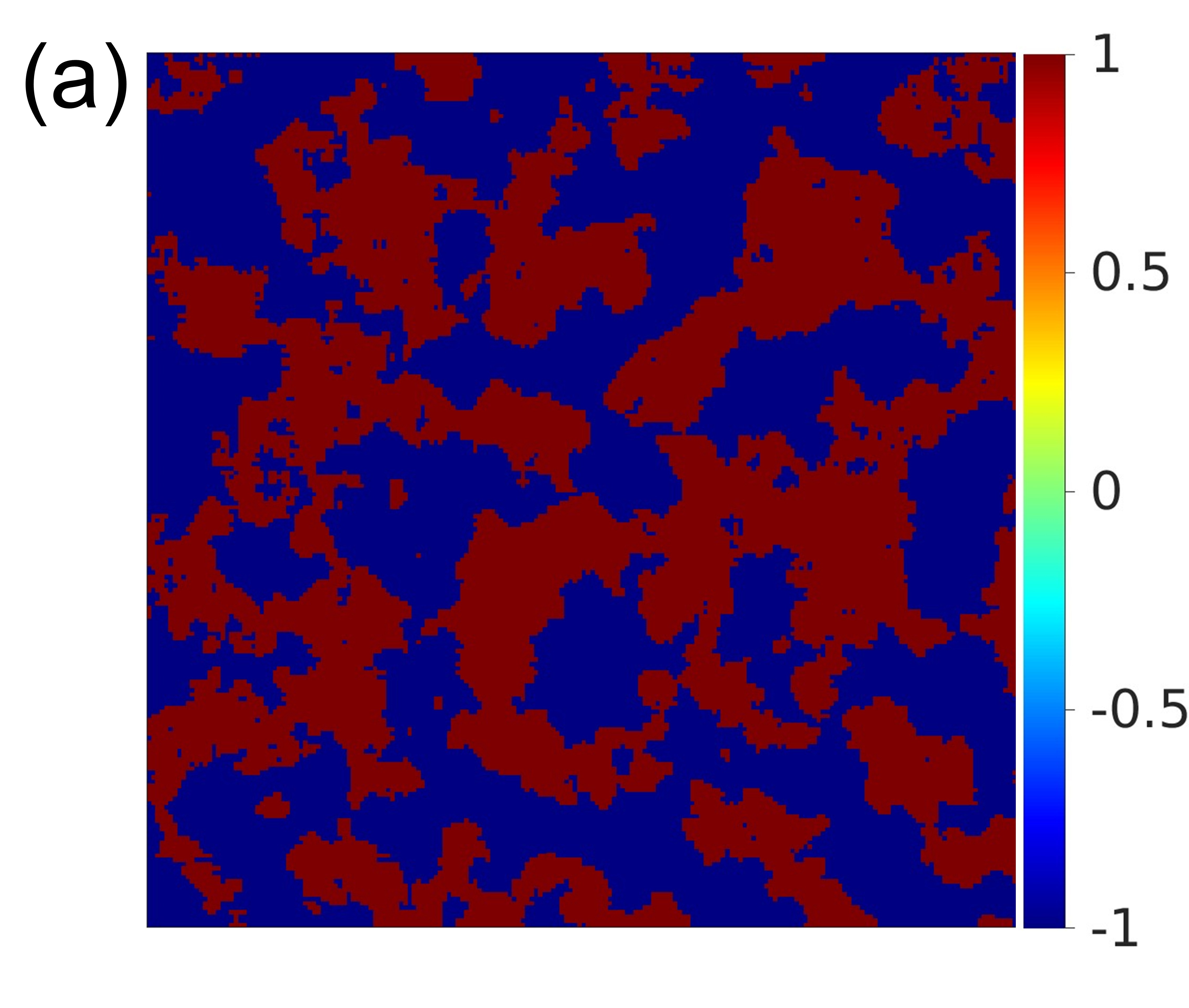} &
        \hspace{-1.18cm} 
        \includegraphics[width=0.32\textwidth,height=0.25\textwidth]{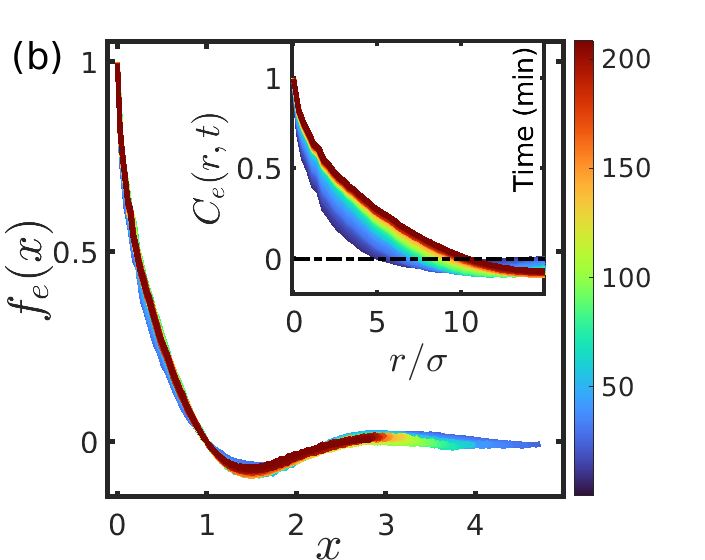} &
        \hspace{-0.7cm} 
        \includegraphics[width=0.33\textwidth,height=0.26\textwidth]{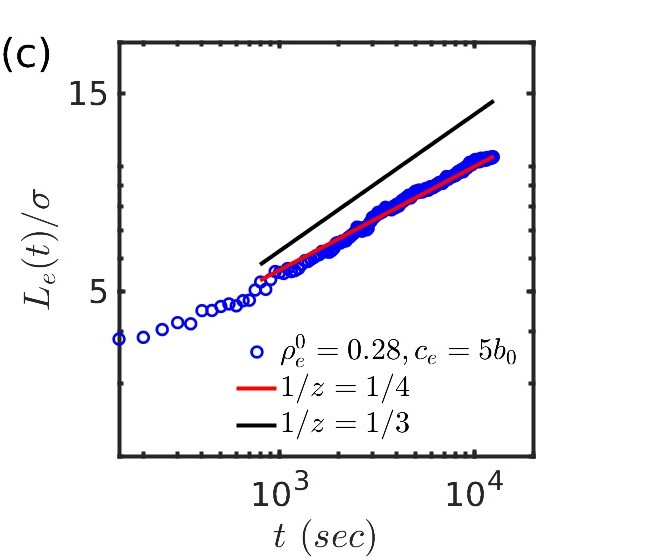} \\
    
        \includegraphics[width=0.39\textwidth,height=0.28\textwidth]{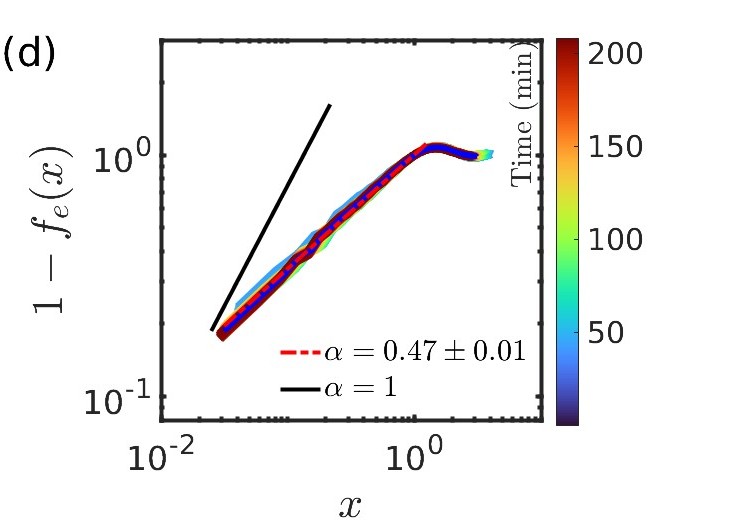} &
        \hspace{-1cm} 
        \includegraphics[width=0.35\textwidth,height=0.27\textwidth]{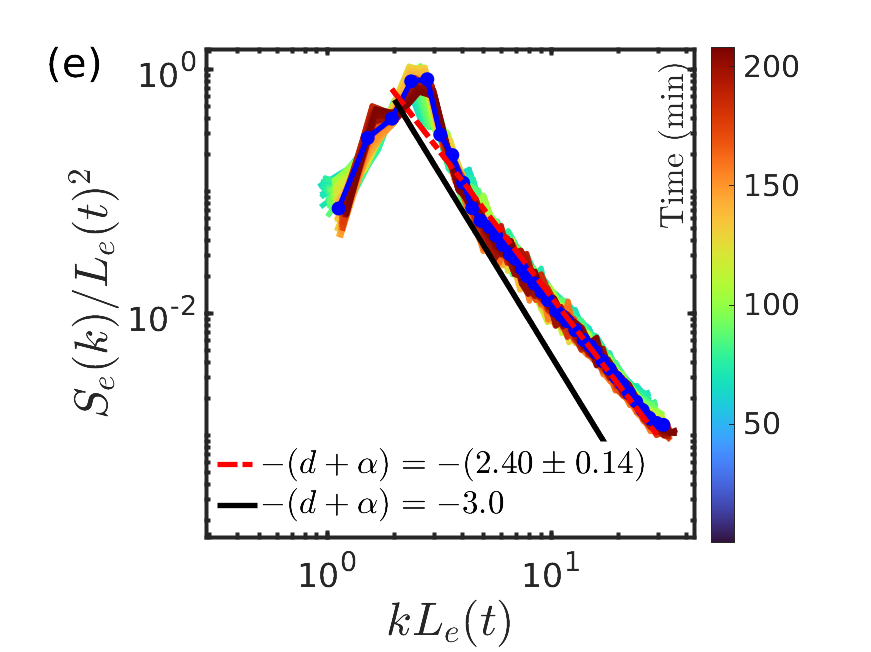} &
        \hspace{-0.5cm} 
        \includegraphics[width=0.3\textwidth,height=0.28\textwidth]{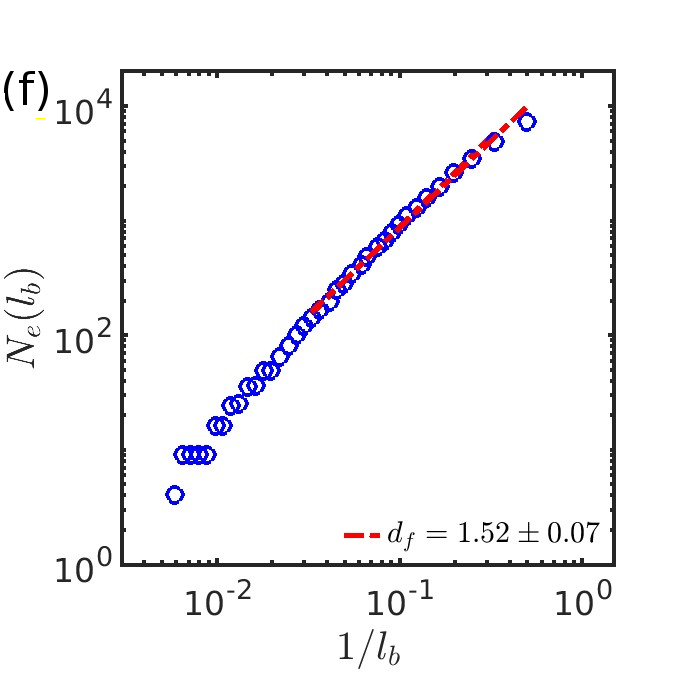}
    \end{tabular}
    
    \caption{Phase ordering of colloids and the nature of interfaces in experiments. 
    (\textbf{a}) A reconstruction of the instantaneous order parameter field, $\phi_e(\mathbf{r},t)$, corresponding to the image presented in Fig.~1(c).
    (\textbf{b}) Main panel: The scaled correlation function represented by $f_e(x)\equiv f(r/L_e(t)) =C_e(r,t)$, where $x=r/L_e(t)$. Inset: The time evolution of the correlation function $C_e(r,t)$ in the transient stages of coarsening. The time lapsed since the start of experiment is color coded. The dashed line depicts $C_e(r,t)=0$. 
    (\textbf{c}) The scaled coarsening length scale $L_e(t)$ is shown as a function of time $t$. The thick red line has a slope of $1/4$, while the slope of the black line is $1/3$. 
    (\textbf{d}) The reduced scaling function $\tilde{f_e}(x)=(1-f_e(x))$ is presented along with lines of slopes $1$ (thick line) and best-fit line (dashed line).
    (\textbf{e}) Scaled structure factor of the order parameter $S_e(k)/L_e(t)^2$ plotted against $kL_e(t)$. The black and red lines correspond to slopes of -3 and best-fit line with slope $-(2.40\pm0.14)$, respectively.
    (\textbf{f}) The count of boxes $N_e(l_b)$ of size $l_b$ is shown against inverse of the box size ($1/l_b$). The red line is a linear fit to the data in the range ($1/l_b=0.0057-0.49$) whose slope is the interfacial fractal dimension, $d_f = 1.52\pm0.07$. Note $l_b$ is normalized by the particle diameter.
    The density of colloids is $\rho_e^0=0.28$ and concentration of bacteria is $c_e=5b_0$.}
    \label{Fig2}
\end{figure}

Next, to investigate cluster growth, their morphology, and interface structure, we introduce an order parameter $\phi_e(\mathbf{r},t)$, which is defined using the local density $\rho_e^l(\mathbf{r},t)$. Following earlier studies \cite{Bray94, Puri09}, the order parameter $\phi_e$ is assigned a value of $+1$ when $\rho_e^l\geq\rho_e^0$ and $-1$ when $\rho_e^l<\rho_e^0$, where $\rho_e^0$ is the average density of colloids. A reconstruction of the instantaneous order parameter displayed in Fig.~\ref{Fig2}(a) corresponds to the image in Fig.~\ref{Fig1}(c). See Supplementary Video SM3 for a visualization of the order parameter evolution, which reveals the presence of strong fluctuations due to the active bath. Following this, we compute two-point spatial correlations of the order parameter $C_e(\mathbf{r},t)$, which is defined as 
\begin{equation}
    C_e(\mathbf{r},t) = \langle \phi_e(\mathbf{r}_1, t) \phi_e(\mathbf{r}_2, t) \rangle - \langle \phi_e(\mathbf{r}_1, t) \rangle \langle \phi_e(\mathbf{r}_2, t) \rangle
    \label{twopointCorr}
\end{equation}
where $\mathbf{r}= \mathbf{r}_1-\mathbf{r}_2$, and the brackets denote the averaging over $\mathbf{r}$. Its Fourier transform is the structure factor $S_e(\mathbf{k})$, which is defined as
\begin{equation}
    S_e(\mathbf{k},t) = \langle \tilde{\phi_e}_{\mathbf{k}}(t)\tilde{\phi_e}_{-\mathbf{k}}(t)\rangle,
    \label{eq:sf}
\end{equation}
where $\tilde{\phi_e}$ is the Fourier transform of $\phi_e$. 
Using isotropy, the correlation function $C_e(\mathbf{r},t)$ and the structure factor $S_e(\mathbf{k},t)$ are averaged over all orientations of $\mathbf{r}$ and $\mathbf{k}$, respectively. The results are shown as functions of the radial distance $r = |\mathbf{r}|$ and the magnitude of the wave vector $k = |\mathbf{k}|$.\\ 

The inset of Fig.~\ref{Fig2}(b) shows $C_e(r,t)$ at increasing times, with the color code indicating the time elapsed in minutes since the start of the experiment. The decay of correlations becomes slower with increasing time, suggesting an increase in the size of clusters. The distance over which the correlations decay to zero is taken as the length scale $L_e(t)$ of the clusters. The correlation function $C_e(r,t)$ displays dynamic scaling, implying the existence of a single characteristic length scale $L_e(t)$. When $C_e(r,t)$ is rescaled by $L_e(t)$, all correlation functions collapse onto a single scaling curve. The scaling function, defined as $f_e(x) = f(r/L_e(t)) = C_e(r/L_e(t))$, is shown in the main panel of Fig.~\ref{Fig2}(b). Next, the growth of the cluster size $L_e(t)$ is shown in Fig.~\ref{Fig2}(c), along with lines of slopes $1/4$ (red) and $1/3$ (black). This unambiguously demonstrates that the cluster size follows $L_e(t)\sim t^{1/z}$, with $z \sim 4$, in our experiments. These results provide clear evidence of deviations from the $t^{1/3}$ LS growth law. Similar results are obtained when particles of larger size ($\sigma=20\mu m$) are used, see section I-C of SI for details. Also, the results remain unchanged when the two-point spatial correlation, $C_e(r,t)$, is computed using the local density as the order parameter, instead of the hardening procedure followed in Eq.\ref{twopointCorr}, see section I-D of the SI. Earlier experiments with active colloids \cite{Granick21, Douchot19} and similar studies \cite{Auradou23} involving colloid-swimmer mixtures reported domain growth consistent with thermal systems, i.e. $L(t) \sim t^{1/3}$. By contrast, our findings are reminiscent of simulation results \cite{Cate14, Puri20}, where domain growth following $t^{1/4}$ was reported. Although there is no consensus on the exact exponent value \cite{Cate14}, all investigations suggest a clear deviation from the conventional exponent of $1/3$. Furthermore, nonequilibrium ordering in lattice models of particles sliding on fluctuating surfaces has also revealed a strong departure from the conventional $t^{1/3}$ behavior. Similar deviations were observed in simulation models of the organization of passive molecules on active cell surfaces \cite{Rao16}. Our results provide strong experimental evidence for slower growth. These features remain robust against variations in the density of colloids and swimmers; see Section I-E of the Supplementary Information.\\

\section{Structure of the interfaces separating the colloid-rich and colloid-poor domains}

We now examine the structure and morphology of the colloidal domains and the interfaces between the colloid-rich and colloid-poor phases. In thermal systems approaching equilibrium, interfacial profiles are governed by free energy minimization, typically resulting in smooth and well-defined interfaces. However, in our system, where the colloids are constantly driven by the swimmers, the interfaces are dynamic due to the continual aggregation and fragmentation of colloidal clusters. A careful examination of the order parameter field, $\phi_e(\mathbf{r},t)$, in Fig.~\ref{Fig2}(a) and the Supplementary Video SM3 clearly reveals interfaces with pronounced roughness and strong fluctuations. A magnified view of the colloidal domains (Fig.~\ref{Fig5}) and the procedure used to identify the interfaces from the order parameter are outlined in Methods. Further insights can be obtained by analyzing the scaled correlation function $f_e(x)$ and the structure factor $S_e(k,t)$. Due to the presence of interfaces, the correlation function $f_e(x)$ exhibits a singularity at short distances ($r/L \rightarrow 0$) and can be expressed as \cite{Majumdar01, Barma00, Puri14}
\begin{equation}
f_e(x)=1-Ax^{\alpha}+..., 
\end{equation} 
where the exponent $\alpha$ encodes the nature of the interfaces. 
For domains with smooth boundaries and no hierarchy in inter-domain separations, $\alpha = 1$, consistent with Porod's law \cite{Glatter82}. To determine the exponent $\alpha$ in our experiments, we analyze the reduced scaling function $\tilde{f}_e(x) = 1 - f_e(x)$, as shown in Fig.~\ref{Fig2}(d). The dashed line represents the best-fit line with a slope $0.47 \pm 0.01$, while the thick line has a slope $1$. The observed cusp exponent, $\alpha = 0.47 \pm 0.01$, points to strong deviations from Porod's law. In particular, the condition $\alpha < 1$ implies the presence of rough interfaces \cite{Puri14}, consistent with the visual impression of Fig.~2(a) and the Supplementary Video SM3. \\

To further substantiate this observation, we present the scaled structure factor in Fig.~\ref{Fig2}(e). In general, the structure factor at large $k$ takes the following form:
\begin{equation}
\tilde{S}_e(k)\sim (k)^{-(d+\alpha)},
\end{equation}
where $d$ is the dimensionality of the system. Theoretically, the exponent is related to the fractal dimension $d_f$ of the interfaces through $\alpha = d - d_f$ \cite{Puri14,Schmidt84,Sorensen01}.  For domains with sharp interfaces, $d_f = d - 1$ and the cusp exponent  $\alpha = 1$. In contrast, the results in Fig.~\ref{Fig2}(e) show a non-integer exponent, $\alpha = 0.40 \pm 0.14$, obtained from a linear fit to the data, indicating rough and diffuse interfaces. The values of $\alpha$ extracted from the reduced correlation function $\tilde{f}_e(x)$ and the scaled structure factor $\tilde{S}_e(k)$ are consistent within experimental uncertainty. The weighted average exponent, $\tilde{\alpha} = 0.47 \pm 0.01$, therefore, corresponds to a fractal dimension $d_f \approx 1.5$. To test this, we directly measured $d_f$ using the box-counting method \cite{Stanley95}, see Methods. Figure~\ref{Fig2}(f) displays the number of boxes $N(l_b)$ of size $l_b$ needed to cover the interface against the inverse box size. Since smaller box sizes that are of the size to the colloids lose information about the interfacial structure in counting, we exclude a few data points from large $1/l_b$ region. A linear fit depicted by the red line yields a slope of $d_f=1.52\pm0.07$, which is close to its expected value. These results demonstrate that the growth and interfacial morphology of colloidal clusters exhibit strong deviations from equilibrium behavior, underscoring the distinct nature of phase separation in active systems. \\

Thus far, we have presented experimental results that highlight the unconventional aspects of phase ordering in colloids dispersed in active liquids. In the following sections, we introduce a coarse-grained model based on a scalar order parameter. Before introducing the model, comments are in order regarding its scope and assumptions. First, hydrodynamic interactions are neglected. Although microswimmers generate hydrodynamic disturbances, several observations indicate that hydrodynamic interactions do not play a dominant role in determining the growth kinetics or interfacial morphology. The quasi-two-dimensional geometry and proximity to solid boundaries strongly screen long-range hydrodynamic interactions \cite{Lauga09,Blake71}. These findings indicate that activity-induced fluctuations, rather than fluid-mediated advection, govern the unconventional phase ordering observed here. Previous studies have reproduced key experimental observations, based on the presence of activity \cite{Palacci23,Granick21}. Second, while microswimmers are known to exhibit chiral trajectories near solid boundaries \cite{Stone06}, our scalar order parameter framework does not account for the effects of chirality \cite{Palacci23, Tanaka24}. Nevertheless, we will show that incorporating only the effects of activity is sufficient to capture essential aspects of the underlying physics. The model reproduces key features of the experimental observations and offers valuable insights into the emergent behavior.

\section{Theoretical framework: Model B with colored noise}

We present a scalar coarse-grained model motivated by our experimental findings. Previous studies had shown that the influence of bacteria on the dynamics of passive colloids can be effectively modeled using temporally correlated noise \cite{Libchaber00}. These models have successfully captured the transport properties of colloids in suspension, particularly at low swimmer densities. The temporal correlations in the noise originate from intrinsic memory, which breaks the time-reversal symmetry in the bacterial dynamics. In addition, the persistent motion of bacteria introduces a characteristic persistence length scale \cite{Hagan25}. When the density of passive colloids is low and their separation greatly exceeds this length scale, interactions between them can be neglected. However, as the separation between colloids approaches the bacterial persistence length, the noise must also be treated as spatially correlated. The range of correlations depends on the bacterial concentration. As the concentration of \textit{E. coli} increases, their motion becomes increasingly coherent, leading to an enhancement of the persistence length. As shown in Table~1 in section I-F of the SI, an estimate of this persistence length can be obtained from the parameter $r_c$, which systematically increases with increasing bacterial concentration.  \\

We therefore model the influence of an active medium on the dynamics of passive colloids as thermal noise correlated in both space and time. This noise is solely due to the finite concentration of bacteria $c$ and vanishes at zero $c$. Thus, the colloidal assembly is modeled as a binary mixture, similar to the mixture of high-density liquid and low-density vapor. Although it is a binary system, our focus is on the segregation dynamics of the passive particles that constitutes a one-component system. The active bath of E. coli remains homogeneous during the segregation process. Therefore, the dynamics of the slow variable, which is the local density field of the colloids $\rho({\bf r}, t) \in [0,1]$, obeys Model B \cite{Halperin77} because the total number of particles is conserved. \\

The corresponding conserved order parameter field is defined as 
$\phi(\boldsymbol{r},t) = 2 \rho(\boldsymbol{r},t) -1 $. The evolution of $\phi(\boldsymbol{r},t)$ is governed by the continuity equation, 
\begin{align}
     &\quad \frac{\partial \phi(\boldsymbol{r},t)}{\partial t} = -\boldsymbol{\nabla} \cdot \boldsymbol{J} +\boldsymbol{\nabla} \cdot (\sqrt{2D(c)}\boldsymbol{\xi}) .
     \label{eq:phi}
\end{align}
Recent experiments have shown that the effective diffusivity of passive colloids varies linearly with the mean concentration of swimmers in the suspension \cite{Libchaber00, Arratia16}, hence we set $D(c) \sim c$. The first term in Eq.~5 is the current $\boldsymbol{J} (\boldsymbol{r},t) = -D(c)\boldsymbol{\nabla}\mu$, 
which is driven by the ``chemical potential'' $\mu$, and the second term denotes the spatio-temporal colored noise of the active bath. The chemical potential $\mu$ is obtained from the functional derivative of the free energy $\mathcal{F}$, $ \mu = \delta \mathcal{F} / \delta \phi$, where the ``free energy functional'' is {expressed in the standard Ginzburg-Landau form \cite{Bray94, Lubensky00}} in dimensionless units as
\begin{equation}
    \mathcal{F}[\phi] = {\int} \left[ -\frac{1}{2} \phi(\boldsymbol{r},t)^2  + \frac{1}{4} \phi(\boldsymbol{r},t)^4 + \frac{1}{2} (\nabla\phi(\boldsymbol{r},t))^2 \right] d \boldsymbol{r}.
    \label{eq:freeenergy}
\end{equation}
Further details of the model are given in section II-A of SI.
Substituting these into Eq.~(\ref{eq:phi}), we obtain the following.
\begin{align}
     \quad \frac{\partial \phi(\boldsymbol{r},t)}{\partial t} = D(c)\nabla^2\left(\frac{\delta \mathcal{F}}{\delta \phi}\right) +\boldsymbol{\nabla} \cdot (\sqrt{2D(c)}\boldsymbol{\xi}) .
     \label{eq:5}
\end{align}
As mentioned earlier, the presence of bacteria will give non-zero $D(c)$ with $D(c = 0) = 0$. We can redefine the time as $t' = D(c)t$ to obtain (dropping the prime)
\begin{align}
     \quad \frac{\partial \phi(\boldsymbol{r},t)}{\partial t} = \nabla^2\left(-\phi + \phi^3 - \nabla^2 \phi\right) +\boldsymbol{\nabla} \cdot \boldsymbol{\xi} .
     \label{eq:5m}
\end{align}
The colored noise $\boldsymbol{\xi}$ is Gaussian in nature with zero mean $\langle \xi_i(t) \rangle = 0$, and the spatial and temporal correlations 
\begin{equation}
\langle \xi_i({\bf r}_1, t_1) \xi_j({\bf r}_2, t_2) \rangle = \delta_{ij} \mathcal{K}_{\lambda \tau}({\bf r}_1 - {\bf r}_2, t_1 - t_2) .
\end{equation}
The correlator $\mathcal{K}_{\lambda \tau}({\bf r}_1 - {\bf r}_2, t_1-t_2)$ is rotationally invariant and has spatial correlation length $\lambda$ and correlation time $\tau$, see section II-B in Supplementary Information for details of the noise. We refer to this model as Model B with colored noise. An estimation of the parameters $\lambda$ and $\tau$ in our experiments is included in section I-F of SI, along with other details. In simulations, we vary these parameters over a comparable range. Starting from a disordered initial condition, equivalent to a homogeneous system, $\phi (\boldsymbol{r},0) = \phi_0 + \mbox{small fluctuations}$, Eq.~(\ref{eq:5m}) is solved numerically as explained in section II-B of the Supplementary Information. The results obtained are described below. Before presenting the results, note that in our experiments the bacteria display chiral motion at the solid–liquid interface near the boundaries. This chirality simply renormalizes the effective diffusivity of the colloids \cite{semwal2024macro, ma2022dynamical}. \\

\begin{figure}[h!]
    \centering
         \raisebox{-2.5mm}{%
         \includegraphics[width=0.43\linewidth]{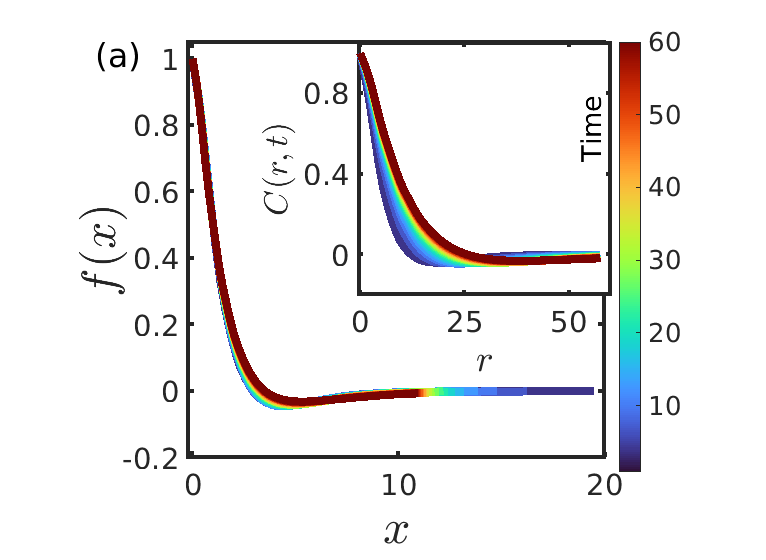}}
    \hspace{-8.5mm}
    \includegraphics[width=0.305\linewidth]{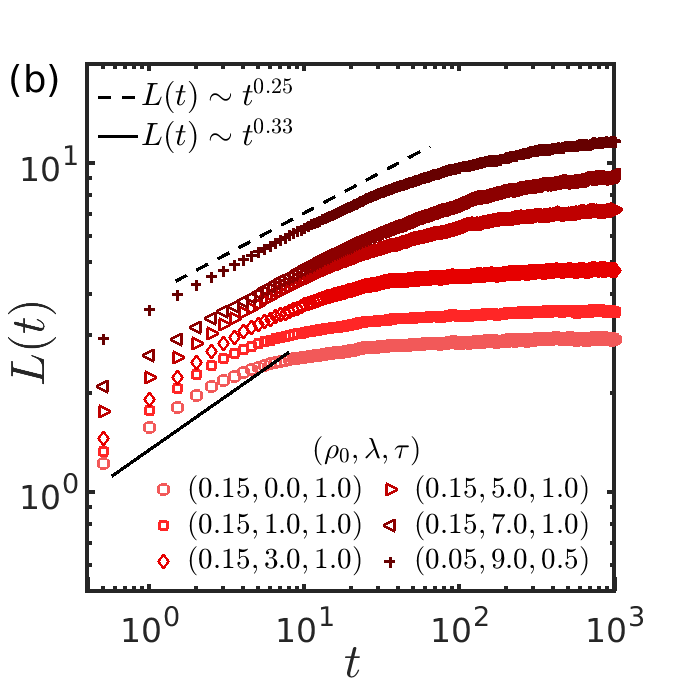}
    \hspace{-4.5mm}
    \includegraphics[width=0.305\linewidth]{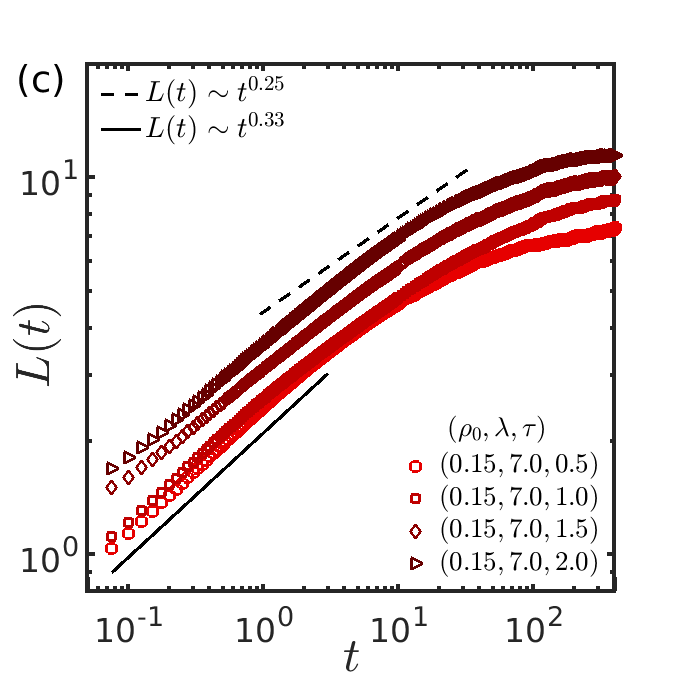}
    \caption{Dynamic scaling of the two-point correlation function, $C(r,t)$, and domain growth kinetics in simulations. 
    \textbf{(a)} Inset: Time evolution of the correlation function. Main panel: Collapse of the correlation functions ($f(x)$) when distances are scaled by the correlation length $L(t)$, denoted by the scaled variable $x = r/L(t)$. These results are obtained using $\rho_0=0.15$, $\tau=1$ and $\lambda=7$. \textbf{(b-c)} The variation of characteristic length $L(t)$ vs. time for different parameters. Data correspond to the range of parameters mentioned in the legends of the plots. The system size under consideration is K = 1024. The slope of thick lines in panels (b-c) is $1/3$ and the dashed line is $1/4$.}
    \label{Fig3}
\end{figure}

\section{Results for Model B with colored noise}

We first characterize the kinetics of domain growth and the evolving domain morphology by calculating the correlation function $C(r,t)$ and the structure factor $S(k,t)$ of the hardened field $\phi(\mathbf{r}, t)$, the same as defined in Eqs. (1) and (2). The hardening procedure maps $\phi > 0$ to $\phi = +1$ and $\phi < 0$ to $\phi = -1$, effectively converting diffuse interfaces into sharp step functions. This approach eliminates artifacts that arise from interfacial thickness. The characteristic length scale $L(t)$ is calculated from the structure factor as the $L(t) = \langle k\rangle^{-1}$, where $\langle k \rangle = \frac{\int k S(k)dk}{\int S(k)d\mathrm{k}}$. The inset of Fig.~\ref{Fig3}(a) shows $C(r,t)$ as a function of $r$ at different times. When the $r$-axis is rescaled by the characteristic length scale $L(t)$, the correlation functions collapse into a single master curve, $f(x)$, as shown in the main panel of Fig.~\ref{Fig3}(a). This collapse demonstrates dynamical scaling, in agreement with our experimental observations. \\


The time evolution of the characteristic domain size, $L(t)$, is shown in Figs.~\ref{Fig3}(b) and \ref{Fig3}(c) for a range of the parameters $\lambda$ and $\tau$. For $\lambda, \tau = 0$, the system shows the usual LS law leading to macrophase separation, see Fig.~S6 (Gaussian noise) in Section II-C of the Supplementary Information and Supplementary Video SM6. For non-zero values of $\lambda$ and $\tau$, we find a sequence of crossovers from $t^{1/3} \rightarrow t^{1/4} \rightarrow t^0$, with two crossover times. This crossover is illustrated in Fig.~\ref{Fig3}(b), where the growth curves represented by $L(t)$ are computed for $\lambda = ${0, 1, 3, 5, 7} at a fixed $\tau=1.0$ and $\lambda =$9 at $\tau = 0.5$. The first crossover time, from $t^{1/3}$ to $t^{1/4}$, strongly depends on $\lambda$. On increasing $\lambda$, the crossover occurs earlier and $t^{1/4}$ growth becomes prominent. It is clear that when $\lambda >3 $, we observe a clear $t^{1/4}$ regime, 
{although it extends over a limited dynamic range because of the subsequent onset of micro-phase separation. This results in an effective growth law of $t^{1/4}$ before the domain size saturates.} 
The variation of $L(t)$ for varying $\tau$ and a fixed $\lambda=7$ is shown in Fig. \ref{Fig3}(c),  we observe $L(t) \sim t^{1/3}$ growth at smaller values of $\tau$ at early times, and a $L(t) \sim t^{1/4}$ growth at intermediate times. Furthermore, in steady state, the system exhibits a strong dynamic state consisting of many interconnected domains that continuously merge and reorganize, while the average domain size in steady state $L_s$ remains approximately constant, as shown in Supplementary Video SM4. As expected from dimensional analysis, the normalised saturation length $L_s/L(0.5)$ shows (a) linear dependence on $\lambda$, and (b) remains invariant with respect to $\tau$. This is shown in Fig.~S7 in section II-D of the supplementary information. The saturation of the domain size together with persistent dynamical restructuring of domains is indicative of a dynamical microphase separated state. \\

We further argue that the saturation of domain size is a genuine signature of microphase separation rather than a finite size effect. This is because the size of the saturated length  $L(t) \sim 10$ in Fig.3(b-c) is much smaller than the system size ($K = 1024$) and remains the same for $K = 1024$ and  $2048$ as shown in Fig.~S8 of section II-E in SI. The emergence of the dynamically arrested microphase-separated state can be understood as follows. The spatiotemporally correlated noise generated by the active bacterial bath creates a continuously fluctuating environment for the phase-separating passive colloids. This active fluctuating background suppresses unrestricted domain growth, leading to dynamical arrest and consequently to microphase separation accompanied by a slower coarsening law. Dynamical arrest in phase-separating binary mixtures has also been reported in turbulent environments \cite{PhysRevLett.112.014502}, where the maximum domain size is limited by the Hinze length, the characteristic scale at which turbulent inertial stresses balance interfacial tension forces. As a result, domains larger than this scale are fragmented by turbulence, preventing further coarsening. Similarly, in our experiments, the characteristic length scale of phase separating colloids is determined by the intrinsic correlation length scale associated with the bacterial bath.\\

To further verify the growth exponent, we analyzed the scaling behavior of the correlation functions $C(r,t)$ plotted against $t^{1/z}$, as shown in Figs.~S9 in section II-F of the Supplementary Information. For $z = 4$, the scaled data show excellent collapse, whereas no such collapse is observed for $z = 3$. This further confirms the slower domain growth in our system, and the results are consistent with our experimental findings in Fig.~2(c). A comparison of the domain growth kinetics in both experiment and simulation is presented in the supplementary videos {SM3 and SM4}, respectively. Both videos reveal passive colloidal domains that remain highly dynamic, with strongly fluctuating interfaces. A detailed analysis of the growth behavior of $L(t)$ for other parameters is provided in section~II-G of the supplementary information. Slower domain growth is observed for all cases and follows $t^{1/4}$ scaling, see Figs.~S10(a)–(i). \\

\begin{figure}[h!]
    \centering
    \hspace{10mm}
    \includegraphics[width=0.39\linewidth]{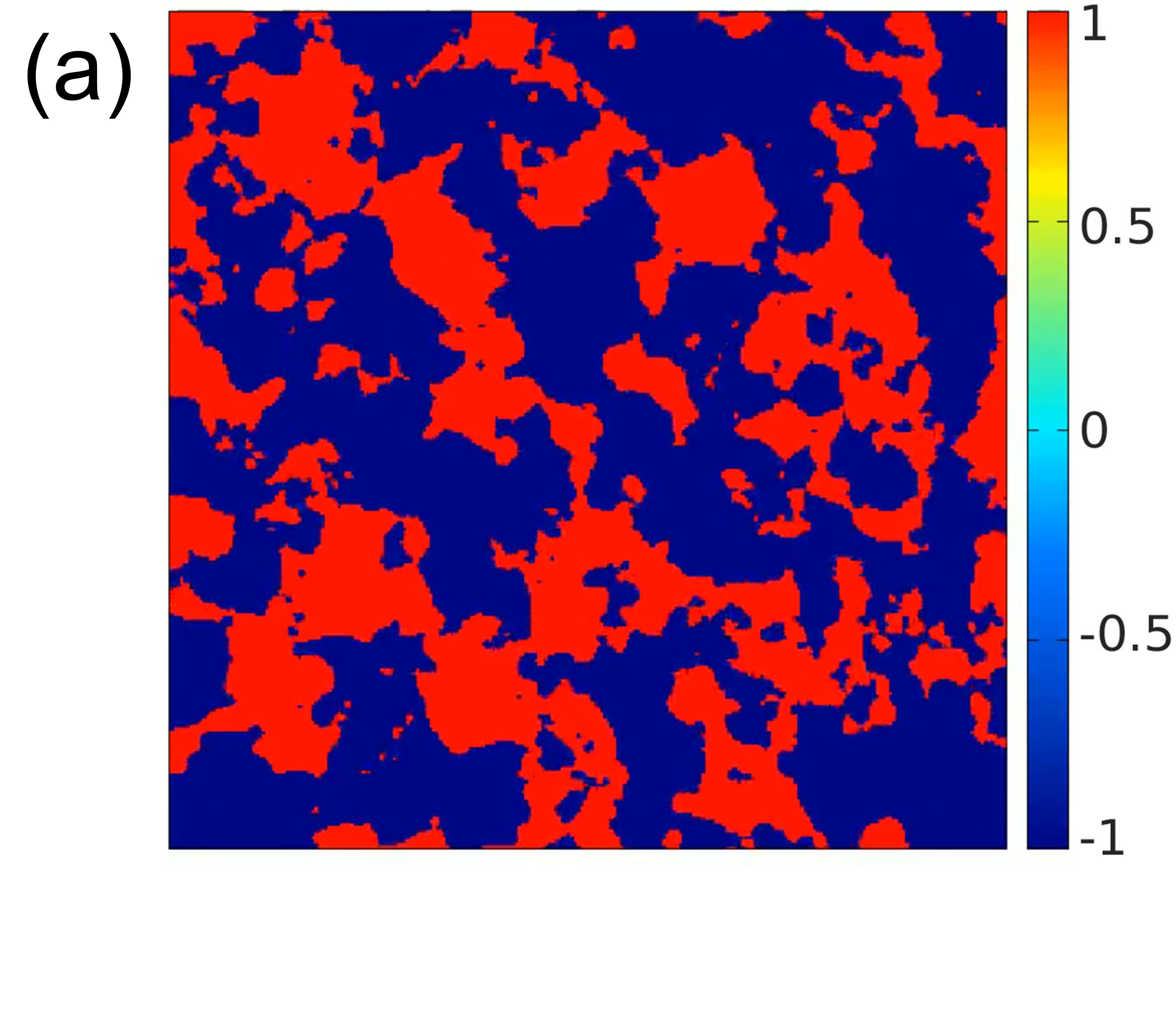}
    \hspace{2mm}
    \includegraphics[width=0.5\linewidth]{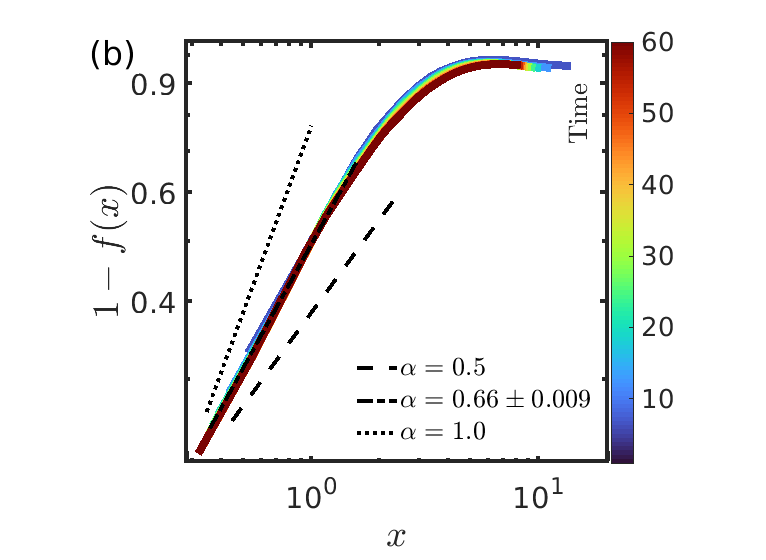}
    \vspace{2mm} \\ 
    \includegraphics[width=0.4\linewidth]{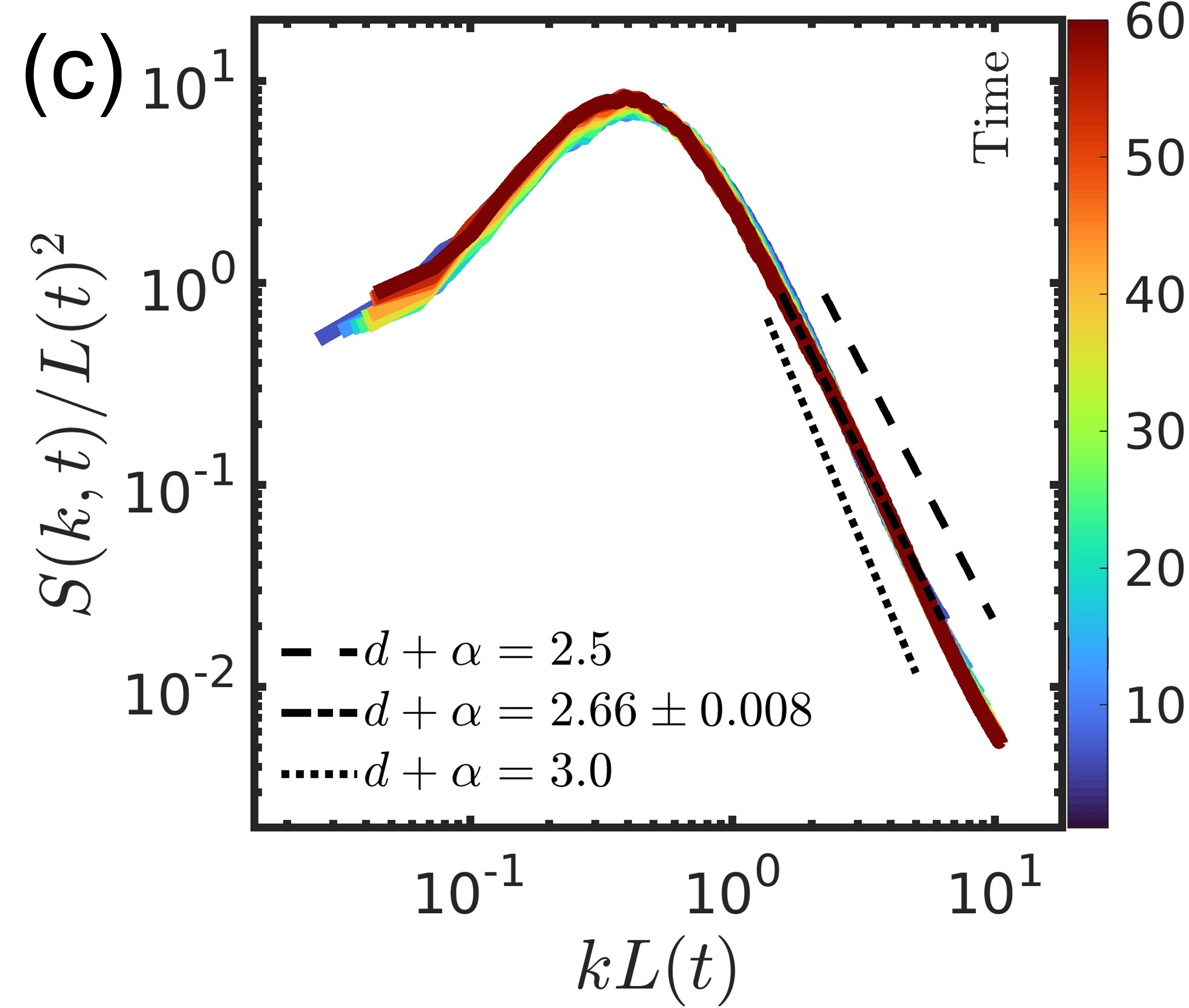}
    \hspace{2mm}
    \raisebox{-1.2mm}{
    \includegraphics[width=0.38\linewidth]{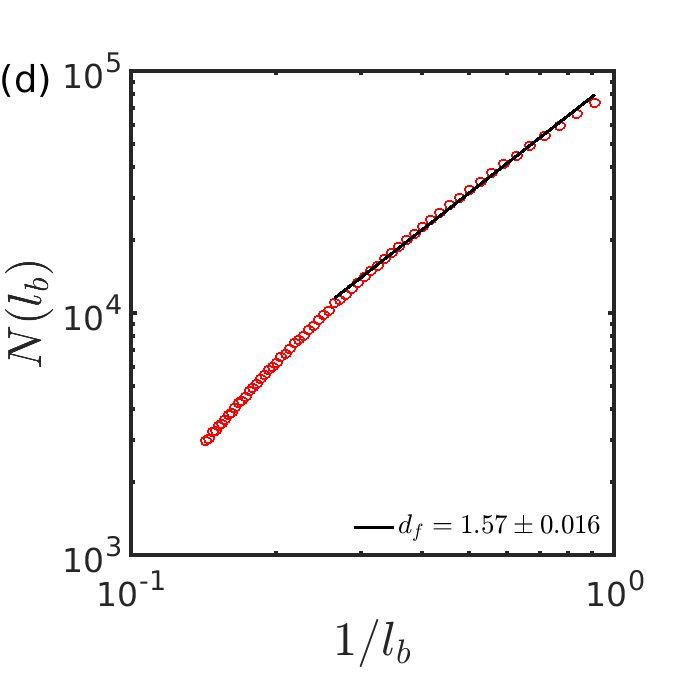}}
    \caption{Nature of interfaces in Model B with colored noise. (\textbf{a}) A snapshot of the hardened order parameter field with the red and blue colors representing the high ($\phi=+1$) and low-density ($\phi=+1$) regions, respectively. 
    (\textbf{b}) The reduced scaling function $\tilde{f}(x)=(1-f(x))$ showing the cusp singularity of the correlation function. Lines of slopes $1$ (dots), $0.5$ (dashes) and a best fit line of slope $0.66$ (dash-dots) are displayed. 
    (\textbf{c}) The scaled structure factor $(S(k))$ for same times (as in panel \textbf{(b)}) illustrating the non-Porod characteristics of the structure factor tail. The color bar shows the time progression. Lines of slopes $3$ (dots), $2.5$ (dashes) and a best fit line of slope $2.66$ (dash-dots) are displayed. 
    (\textbf{d}) The number of boundary points $N(l_b)$ vs. $1/l_b$, where $l_b$ is calculated with the increment of $0.1 \times \Delta x$. The thick line ($1/l_b$ between (0.2 - 0.9)) represents the power law fit with slope $d_f = 1.57\pm0.016$ which represents the fractal dimension of the system. 
    The plots are generated for system size $K = 1024$ with $(\rho_0, \lambda, \tau ) = (0.15, 7.0, 1.0)$.}
    \label{fig:intcr}
\end{figure}

Next, we investigate the nature of domain walls in our model. A reconstruction of the order parameter, $\phi(\mathbf{r},t)$, shown in Fig.~\ref{fig:intcr}(a) reveals similarities with the experimental image in Fig.~\ref{Fig2}(a). The domains are dynamic, and the roughness of interfaces is apparent. These similarities in the structure of the interfaces can be further quantified by computing the short-distance behavior of $C(r,t)$ in Eq.~(3). In Fourier space, $S(k,t)$ at large $k$ is given by Eq.~(4) \cite{Puri14, Puri14-2}. The reduced scaling function $1-f(x)$ is presented in Fig.~\ref{fig:intcr}(b) and the scaled structure factor $L^{-d}S(k,t)$ is shown in Fig.~\ref{fig:intcr}(c). 
{The data in both figures exhibit a clear scaling collapse at all times, yielding $\alpha \simeq 0.66$, which differs significantly from the experimentally measured value of $\alpha \simeq 0.47$. This discrepancy suggests that the interfaces are rougher in the experiments than in the simulations.}
It could be a consequence of the coarse-grained model that averages out the microscopic roughness present in experiments. The cusp exponent is robust and shows little variation for other values of $\lambda$; see section~II-G of the Supplementary Information for details. The interfaces in our simulations were identified using the order parameter of the lattice points. We followed the same approach as outlined for experiments; see Methods for details. According to theoretical models \cite{Puri14,Schmidt84,Sorensen01}, the fractal dimension of the interfaces is $d_f = d-\alpha \simeq 1.34$. The box counting method was followed to calculate the fractal dimension directly from the interfaces found in our simulations. Figure~\ref{fig:intcr}(d) shows the number of boundary points $N(l_b)$ vs. $1/l_b$. Similar to experiments, to obtain fractal dimension from $N(l_b)$ vs. $1/l_b$, we exclude a few data points from large $1/l_b$ regime. A linear fit depicted by the red line yields a slope of $d_f = 1.57 \pm 0.016$. {The value of $d_f$ estimated from Fig.~\ref{fig:intcr}(d) is sensitive to the range of data considered for the fitting, which in the present analysis corresponds to $\frac{1}{l_b}=0.2$--$0.9$.} \\

All together, the presence of colored noise results in a steady state that is highly dynamic and is characterized by continuous fragmentation and aggregation of colloidal domains. This is accompanied by persistent interfacial restructuring and a possible implication of this dynamic state is the slowing down of domain growth processes. In addition, the interfacial properties significantly deviate from their equilibrium counterparts.\\

\section{Discussion and Conclusions}

In summary, we investigated the coarsening dynamics of colloidal clusters in active liquids using a combination of experiments and a coarse-grained theoretical framework. We find that homogeneous mixtures of colloids and active swimmers are inherently unstable, leading to spontaneous phase separation through the emergence of dynamic clusters. The kinetics of these phase-separating domains raises an important question: Does the system evolve toward microphase separation or toward macrophase separation? In the case of microphase separation, the characteristic domain size is expected to saturate to a steady-state value. This is indeed observed in our simulations on longer timescales. The growth of the domains is arrested and the length scale saturates after an initial growth. However, such a saturation can also arise when the domain size becomes a significant fraction of the system size or when the finite colloid mass limits further coarsening. We have verified that it is not a finite-size artifact as evident from the finite size analysis presented in section II-E of Supplementary Information. Unlike the usual microphase separation, the domains observed in our experiment as well as in theoretical model are very dynamic, as can be seen in the Supplementary Videos {SM3-SM5}. Our experiments capture the early stages of coarsening reliably, however, the late-time regime extends beyond the accessible experimental timescales. What is apparent is that the structures observed in the experiments are highly dynamic, with domains exhibiting pronounced temporal fluctuations (Supplementary Video {SM3}). This microphase separation has been reported in other field-theoretic models, such as AMB+ \cite{pradeeparxiv}, and in systems that undergo noise-induced phase segregation \cite{Pagonabarraga24}. \\

Our study uncovers several novel effects of activity on the phase ordering of colloids. The correlation functions of the scalar order parameter exhibit dynamic scaling, with the characteristic length scale exhibiting a slower growth following $L(t) \sim t^{1/4}$ scaling. While such deviations from classical thermal behavior have been reported in several numerical studies \cite{Puri21, Cate14, Rao16}, experimental investigations involving active colloids and colloid–bacteria mixtures have generally observed Ostwald ripening with $t^{1/3}$ growth \cite{Granick21, Douchot19, Auradou23}. To elucidate the anomalous coarsening behavior observed in our system, we extended the standard Model B framework for conserved order parameters by introducing spatio-temporally correlated (colored) noise. The correlation length and time scale of this noise are tuned to reflect the properties of the active medium. The resulting growth kinetics, with $L(t) \sim t^{1/4}$, are in excellent agreement with our experimental observations and exhibit a marked departure from the classical Lifshitz–Slyozov (LS) law. {This $t^{1/4}$ growth regime in our simulations becomes prominent at larger spatial correlation lengths $\lambda$. However, this regime extends over a limited dynamic range because activity subsequently arrests coarsening, leading to saturation of the characteristic domain size and microphase separation. Thus, the observed $t^{1/4}$ behavior should be interpreted as an effective growth regime preceding dynamic arrest, rather than as an asymptotic growth law}. Moreover, the correlation functions display a cusp singularity, indicative of rough interfaces separating colloid-rich and colloid-poor domains. This is further supported by the structure factor, which reveals non-Porod behavior in both experiments and theory. A direct estimate of the fractal dimension confirms the irregular and diffuse character of the interfaces. \\

The presence of fractal interfaces and the deviation of domain growth from the standard LSW law arise from the active nature of the background bath, which breaks time-reversal symmetry and renders the system intrinsically far from equilibrium. The steady state remains dynamically evolving, with continuous coalescence and breakup of dense colloidal clusters, accompanied by persistent interfacial restructuring (see SM3 and SM5). Further, our simulations reveal that the initial growth of the domains is arrested on longer timescales, leading to fluctuation dominated microphase separation. Together, these findings demonstrate that activity fundamentally modifies the coarsening dynamics, giving rise to a new class of phase-ordering behavior in colloids suspended in active fluids. \\

The  {effective $t^{1/4}$ growth regime} \sout{1/4 growth exponent} and fractal interfaces are not unique to colored noise and can arise in several passive and active systems through different mechanisms. For example, passive systems with order-parameter–dependent mobility exhibit $t^{1/4}$ growth due to surface diffusion, while active systems such as Active Model B and Active Model B+ show similar scaling due to rotational flows at domain boundaries [8–10]. Fractal interfaces and dynamic arrest have also been reported in systems with quenched disorder and in block copolymer phase separation [11–14]. Nevertheless, Model B with colored noise provides a natural and physically motivated framework for describing phase separation in colloid–microswimmer systems, capturing the key experimental observations of slow domain growth, dynamic arrest, and complex interfacial morphology. \\
 
Our study has relevance to biological systems, where phase separation occurs in actively driven environments such as the cytoplasm and nucleus, and the ATP-dependent processes continuously inject energy \cite{Rosen17,Hyman11}. We provide an experimentally grounded mechanism by which activity, modeled as spatio-temporally correlated noise, modifies coarsening kinetics and interfacial structure, giving rise to fluctuating interfaces and dynamic microphase separation. Some of these features closely parallel those of biomolecular condensates \cite{Alberti25}, which often exhibit arrested growth and continual remodeling, suggesting a physical route by which living systems maintain mesoscale organization without macroscopic demixing. More broadly, activity emerges as a powerful control parameter in soft and engineered materials, enabling tunable domain growth laws, interfacial geometry, and steady-state length scales \cite{Dogic22-1}. By generating interfaces with enhanced fluctuations and non-integer fractal dimensions, active baths can strongly influence mechanical response, transport, and permeability, offering new strategies for stabilizing heterogeneous materials and controlling morphology without external forcing. The colored-noise extension of Model B introduced here provides a coarse-grained framework for linking microscopic activity to macroscopic material behavior, establishing active–passive mixtures as a versatile platform for nonequilibrium self-organization across biological and soft matter systems. Our study lays the foundation for the study of many other interesting phenomena, including melting of passive colloidal crystals and glass transition in active baths.\\

\section*{Methods}

\subsection{Sample preparation}

The suspension of bacteria was realized using motile E. coli strain U5/41 purchased from the National Center for Microbial Resources (NCMR), Pune. The bacteria were grown using Luria-Bertani (LB) broth as growth medium and later suspended in a custom-made minimal motility (MM) medium for sustained motility without division. \\

Cells were cultured following well-established protocols in the literature \cite{Adler67,Lauga15}. E. coli cells (U5/41 type strain) stored at 4$^\circ C$ as glycerol stock, were grown overnight at $37^{\circ}$C on an LB agar plate containing 1$\%$ tryptone, 1$\%$ NaCl, 0.5$\%$ yeast extract and 1.5$\%$ agar. A single colony was then added to 10 $ml$ of LB broth and incubated at $37^{\circ}$C and 210 rpm until the OD$_{600}$ (optical density at 600 nm wavelength) reached 1.3. The bacterial cells were harvested and washed three times with minimal motility (MM) medium by centrifugation at 3000 rpm for 5 minutes at room temperature to remove any residue of LB broth. The pellet was then resuspended in MM medium to achieve the desired concentrations. The concentration of bacteria in most of the measurements is $c_e=5~b_0$, where $b_0 = 6 \times 10^9$ cells/ml. \\

The motility medium contains a buffer to maintain a pH of 7.0 (a combination of K$_2$HPO$_4$ and KH$_2$PO$_4$), a chelating agent to protect motility from inhibition by heavy metal traces (EDTA) and an energy source (L-serine)\cite{Adler66}. Thus, the composition of the custom-built minimal motility (MM) medium is $10mM$ potassium phosphate (pH 7.0), $0.1mM$ EDTA, $0.002\%$ Tween-20, and $50~mM$ L-Serine. In these mediums, bacteria remain motile without dividing. \\

\subsection{Fabrication of sample chamber for imaging colloids in active liquids}

Polystyrene particles of sizes $15~\mu m$ procured from MicroParticles, GmBH, Germany, were used as passive beads. The colloids are sulfate-functionalized polystyrene particles and carry a negative surface charge. The solvent also contains motility medium with dissolved salts that strongly screen the electrostatic double-layer interactions. Generally, the corresponding Debye length is only a few nanometers, much smaller than the particle diameter ($\sigma=15\mu m$). Consequently, the colloid–colloid interactions are nearly hard-sphere-like with a weak residual electrostatic repulsion. However, the phase separation reported here is therefore not driven by intrinsic colloidal attractions but by activity-mediated effective interactions generated by the bacterial bath.\\

To prevent the beads from sticking to the coverslips, we coated them with PEG-8000 \cite{Pucadyil17}. The observation chamber was created using a double-sided tape with a circular cavity of $1cm$ diameter and $100\mu m$ depth, which was glued to a PEG-coated coverslip. A sketch of the sample chamber is shown in Fig.~S1 in SI. The colloids and bacteria were thoroughly mixed and then filled in the sample chamber.

\subsection{Data acquisition, image analysis and interface detection}

A Leica DMi8 inverted microscope was used to acquire images of the sample. The images were captured in bright-field mode using a 5× objective and a Basler acA2040-180km camera with a resolution of 2048 × 2048 pixels and a sensor size of 11.3 mm × 11.3 mm, at a frame rate of 2 FPS over a total duration of 210 minutes. For image analysis, we used the open-source Trackpy application \cite{Trackpy} to determine the positions of colloidal particles. Trackpy implements the Crocker and Grier algorithm \cite{Grier96}, a robust method to locate colloidal particles in microscopy images by identifying bright features against a darker background.\\

Once the particle centers were identified, the coarse-grained local density $\rho_e^l$ was calculated using boxes of size $2\sigma \times 2\sigma$. Since images are of size $2048\times2048$ pixels and the size of a particle is 13.68 pixels, the size of the coarse-graining grid cell is nearly $27\times27$ pixels. The order parameter $\phi_e$ in our experiments depends on the local density $\rho_e^l$. Specifically, $\phi_e$ is assigned a value of $+1$ when $\rho_e^l \geq \rho_e^0$ and $-1$ when $\rho_e^l < \rho_e^0$.\\

\begin{figure}[h!]  
    \centering
    \includegraphics[width=0.43\linewidth]{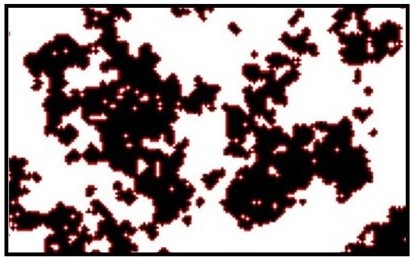}
    \caption{A magnified view of the colloidal clusters. Black and white colors indicate colloid-rich and colloid-poor phases, respectively. The interfaces are marked in red color.}
    \label{Fig5}
\end{figure}

The order parameter $\phi$ is thus a binary variable: $\phi = +1$ indicates the colloid-rich phase, while $\phi = -1$ corresponds to the colloid-poor phase. Each lattice point is further categorized according to the number of $+1$ neighbors, denoted by $n_n$. Points with $n_n = 4$ are considered core points, located within clusters. Points with $2 \leq n_n \leq 3$ that have at least one neighboring core point are classified as edge points, residing at the interface. Points that do not satisfy these conditions are considered isolated and are located in low-density regions. An illustration of this procedure is shown in Fig.~\ref{Fig5}, where the core points are colored black, low-density regions are colored white, and the interfacial points are colored red.\\

\subsection{Calculation of fractal dimension of the interface}

The box counting technique was used to estimate the fractal dimension of the interface. The first step was to locate the interface that separates the colloid-rich and colloid-poor phases using the procedure outlined in the previous paragraph. The box counting method was then applied to determine the fractal dimension. In this approach, a grid of square boxes with side length $l_b$ was overlayed on the binary image, and the number of boxes $N(l_b)$ that intersect the interface was counted. This procedure was repeated for varying box sizes, ranging from $2\sigma$ to $L_{box}/2$, where $L_{box}$ is the length of the field of view in the experiment and the box size in the simulations. Note that $l_b$ is expressed in terms of the diameter of the particle. The interfacial fractal dimension ($d_f$) was then obtained using the following relation:
\begin{equation*}
    d_f = \lim_{l_b\to0} \frac{\log N(l_b)}{\log (1/l_b)}  
\end{equation*}
This method allowed us to quantify the complexity of the interface on different length scales.

\section*{Acknowledgements}
The authors thank Dibyendu Das, Mustansir Barma, and Kabir Ramola for enlightening discussions. P.K. and V.C. thank Sayan Maity for help in preparing a supplementary figure. V.C. acknowledges financial support from IISER Pune and DST/SERB under the project grant CRG/2021/007824 and MHRD for Stars grant MoE-STARS/STARS-2/2023-0909. P.J. acknowledges the DST INSPIRE fellowship for funding this project and P.S.M. gratefully acknowledges UGC for research fellowship. The support and resources provided by the PARAM Shivay Facility under the National Supercomputing Mission, Government of India, at the Indian Institute of Technology, Varanasi, are gratefully acknowledged by all authors. S M thanks DST-SERB (ANRF) India, CRG/2021/006945 and MTR/2021/000438 for financial support. P. J.,  P.S.M., and S.M. also thank the Center for Computing and Information Services at IIT (BHU), Varanasi.

\medskip
%

\clearpage

\setcounter{figure}{0}
\renewcommand{\thefigure}{S\arabic{figure}}
\setcounter{table}{0}
\renewcommand{\thetable}{\arabic{table}}
\setcounter{section}{0}
\renewcommand{\thesection}{\arabic{section}}




\title{
Supplementary Information for \\
Unconventional Growth Kinetics and Fractal Interfaces of Colloidal Phase Separation in Active Liquids \\
}
\author{Pragya Kushwaha$^{1}$}
\author{Pratikshya Jena${^2}$} 
\author{Partha Sarathi Mondal${^2}$}
\author{Nayanthara J R $^{1}$}
\author{Sanjay Puri$^3$} 
\author{Shradha Mishra${^2}$} 
\author{Vijayakumar Chikkadi$^{1}$}
\thanks{Corresponding authors : purijnu@gmail.com, smishra.phy@itbhu.ac.in, vijayck@iiserpune.ac.in}

\affiliation{$^{1}$ Indian Institute of Science Education and Research Pune, India 411008}
\affiliation{$^{2}$ Indian Institute of Technology (BHU) Varanasi, India 221005}
\affiliation{$^{3}$ School of Physical Sciences, Jawaharlal Nehru University, New Delhi, India 110067}

\maketitle
\newpage

\section{Experiments}

\subsection{Experimental set-up}

\begin{figure}[h]
    \centering
    \includegraphics[width=0.5\textwidth]{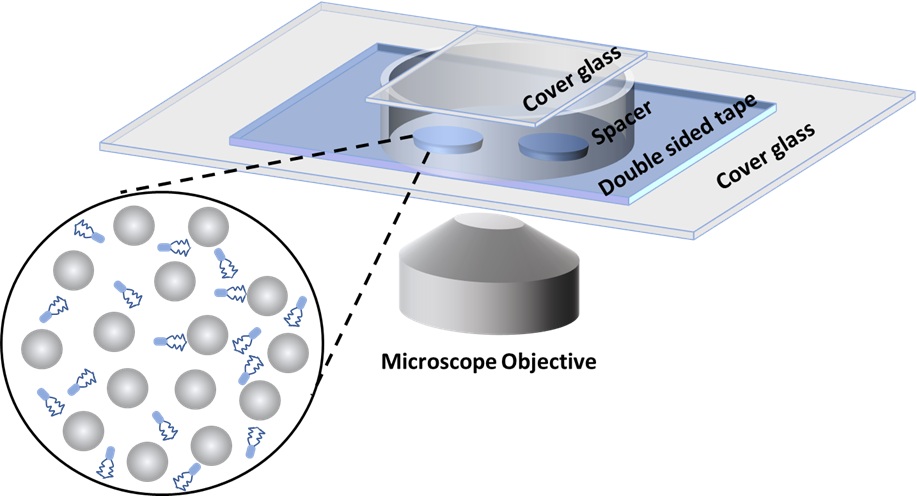}
    \caption{Sketch of the sample chamber used in our experiments. The dimensions of the system are not to scale. The double-sided tape has a thickness of $100\mu m$ and it has two circular cavities. One of them is filled with the suspension of colloidal particles and E. coli, and the other cavity contains DI water. The cover glass on top of the spacer minimizes the evaporation losses.}
    \label{sample}
\end{figure}

The observation chamber was constructed using double-sided tape with a circular cavity of $1cm$ diameter and $100\mu m$ depth, which was glued to a PEG-coated coverlip. A sketch of the sample chamber is shown in Fig. \ref{sample}. The dimensions of the system are not to scale. We used double-sided tape with two cavities. One of the cavities was filled with suspension of colloidal particles and E. coli, while the other contains deionized water. The cover glass on top of the spacer minimized evaporation losses and had small openings in two corners to prevent oxygen depletion, as shown in the schematic diagram.

\begin{figure}[h!]
    \centering
    \includegraphics[width=0.4\textwidth]{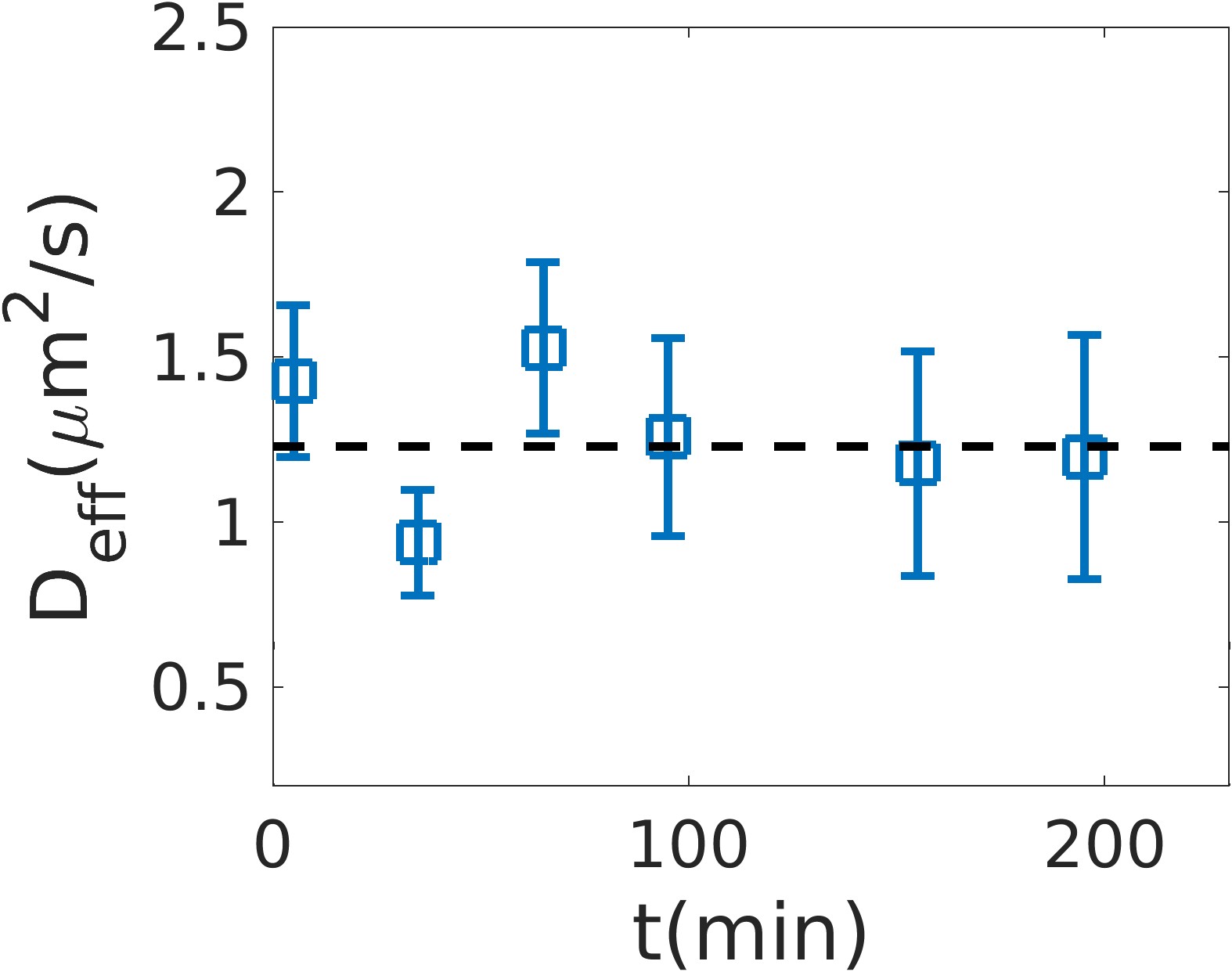}
    \caption{The effective diffusion constant $D_{\textnormal {eff}}$ of colloidal particles in active liquids at a bacteria concentration $c_e=5b_0$. The area fraction of colloids is $\rho_e^0=0.005$. The dashed line indicates the mean value. The error bars are obtained from the standard deviations of $D_{\textnormal {eff}}$. }
    \label{Deff}
\end{figure}

\subsection{Constancy of effective diffusion coefficient in experiments}

As our experiments are conducted over a maximum duration of $210-250$ minutes, we calculated the effective diffusion constant $D_{\text{eff}}$ to ensure that the bacterial activity remains constant over this duration. The effective diffusion constant $D_{\text{eff}}$ in our study is calculated using the expression \cite{Libchaber00}
$$D_{\text{eff}} = \frac{1}{4} \frac{\delta \langle\Delta r^2(t)\rangle}{\delta t}.$$ 
The $D_{\text{eff}}$ is determined at different times starting from the beginning of the experiment and the results are shown in Fig.~\ref{Deff} at dilute densities of colloidal particles corresponding to an area fraction $\rho_e^0= 0.005$ and a swimmer concentration of $c_e = 5b_0$. The estimate of $D_{\text{eff}}$ fluctuates around a mean value, suggesting the constancy of swimmer activity over the duration of the experiment.

\subsection{Coarsening of colloids at various size ratios of colloids and swimmers}

In a previous study \cite{Chikkadi23}, some of the current authors investigated the effect of the size ratio of colloids and microswimmers on the effective interactions between colloids in active liquids of E. coli. Decreasing the size ratio was shown to give rise to smaller clusters of colloids because of weaker effective interactions. We expect the unconventional features of rough interfaces and a slower growth exponent to persist at larger size ratios. \\

\begin{figure}[h!]
    \centering
    \includegraphics[width=0.4\linewidth]{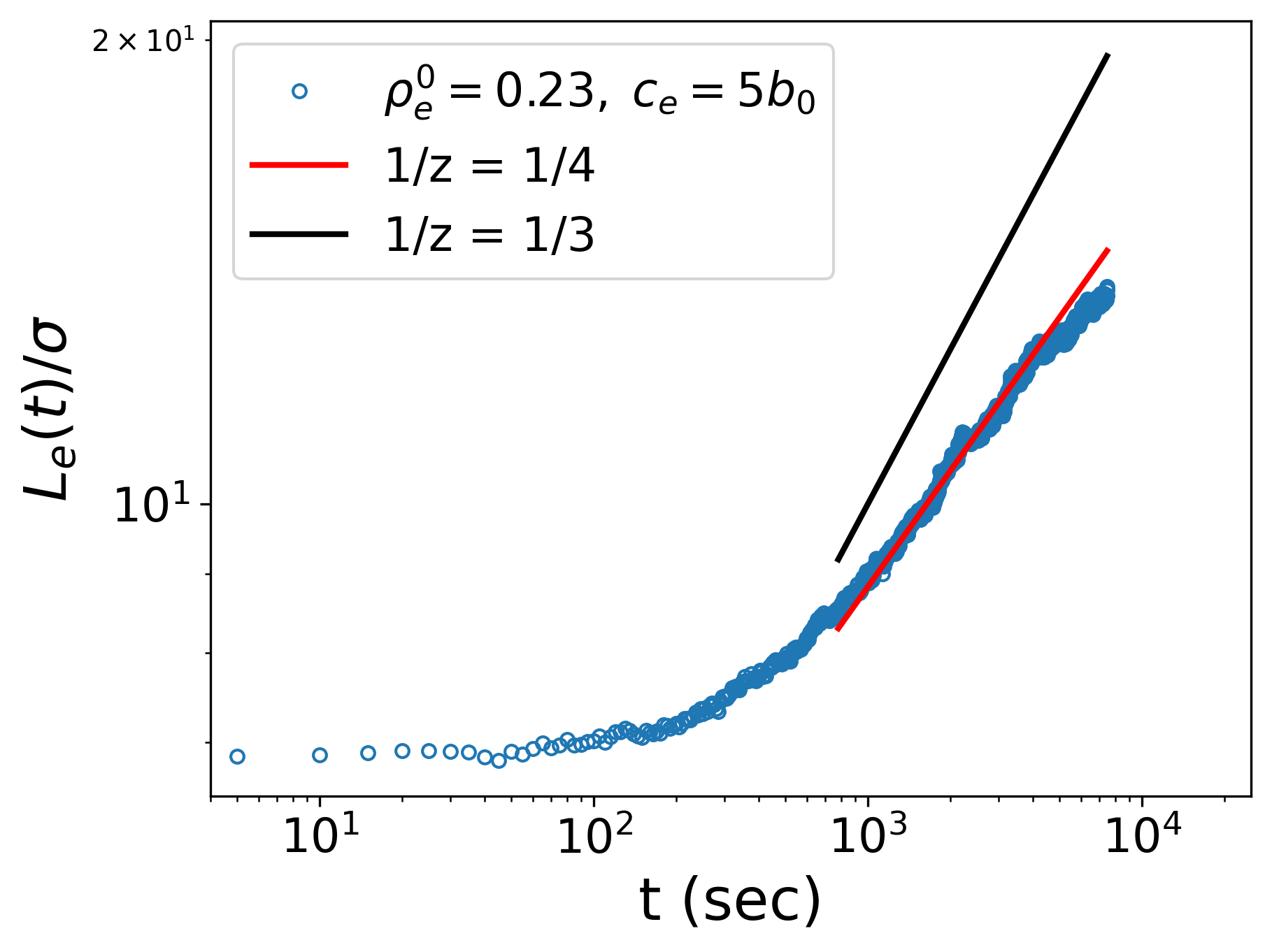}
    \caption{The scaled coarsening length scale $L_e(t)$ is shown as a function of time t. The red line has a slope of 1/4, while the slope of black line is 1/3.}
    \label{Fig-PS20}
\end{figure}

We performed additional experiments to demonstrate these results. The supplementary video SM2 shows $10\mu m$ colloidal particles in a bacterial bath of concentration $c_e=5b_0$, which shows very small clusters. However, colloidal particles of size $20\mu m$ display strong dynamic clustering with $L_e(t)\sim t^{1/z}$, where $z\sim4$, as shown in Fig.~\ref{Fig-PS20}. The experiments reported in the main manuscript used colloids of size $15\mu m$. \\

\subsection{Effect of hardening of the order parameter}

\begin{figure}[h!]
    \centering
    \includegraphics[width=0.5\linewidth]{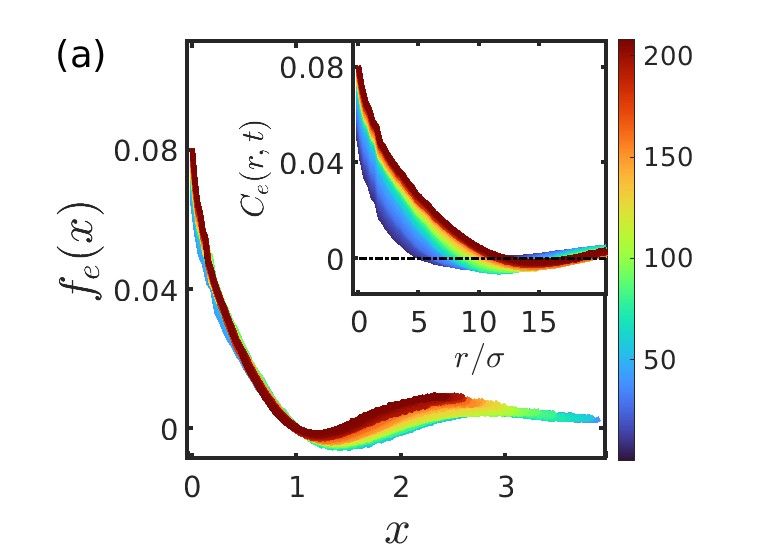}
    \hspace{0.5cm}
    \includegraphics[width=0.45\linewidth]{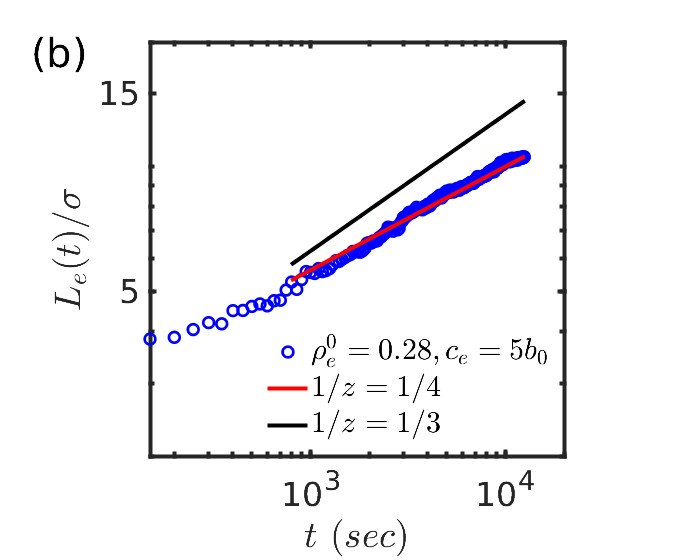}
    \caption{Panel (a) shows the scaled correlation function, $f_e(x)$, while the inset displays the corresponding correlation function, $C_e(r,t)$. Panel (b) presents the characteristic length scale, $L_e(t)$. All quantities are computed using the local density field as the order parameter.}
    \label{Fig-hardening}
\end{figure}

The order parameter, $\phi_e(\mathbf{r},t)$, in our study is defined using the local density $\rho_e^l(\mathbf{r},t)$. Following earlier studies \cite{Bray94, Puri09}, the order parameter $\phi_e$ is assigned a value of $+1$ when $\rho_e^l\geq\rho_e^0$ and $-1$ when $\rho_e^l<\rho_e^0$, where $\rho_e^0$ is the average density of the colloids. This \textit{hardening} procedure is commonly used in the domain growth literature to elucidate the Porod law \cite{Puri88,Oono88}. Our findings are not affected by this mapping. In the Fig.~\ref{Fig-hardening}, the experimental results are computed without hardening the order parameter. The local density is considered as the order parameter. These results do not show any major difference from those presented in Fig.~2 of the manuscript, which are computed using a hardened order parameter. \\

\subsection{Robustness of coarsening exponent at varying densities of colloids and swimmers}

\begin{figure}[h]
    \centering
    \includegraphics[width=0.32\textwidth]{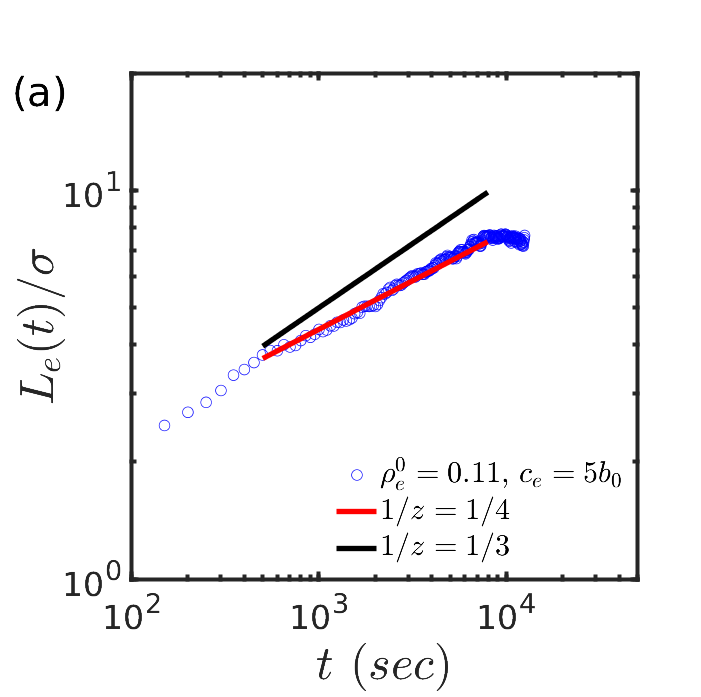}
    \includegraphics[width=0.32\textwidth]{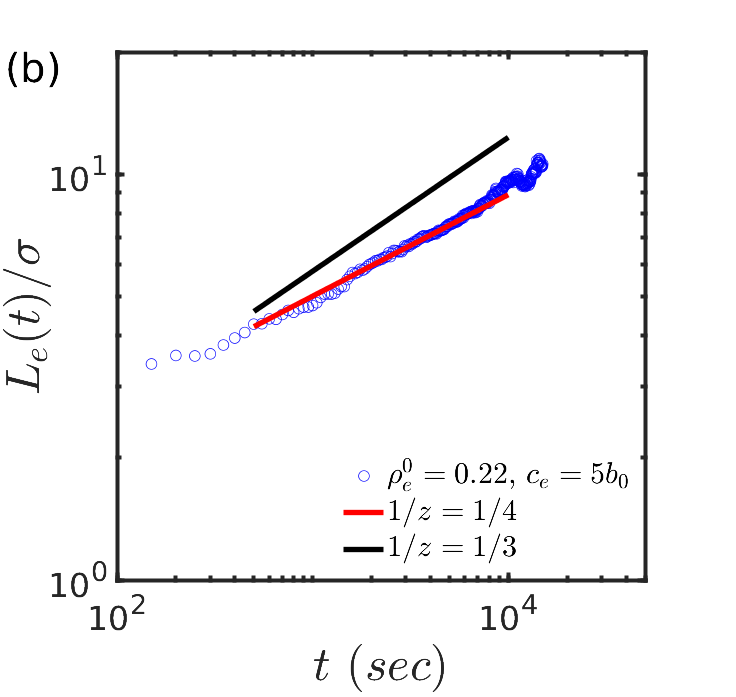}
    \includegraphics[width=0.32\textwidth]{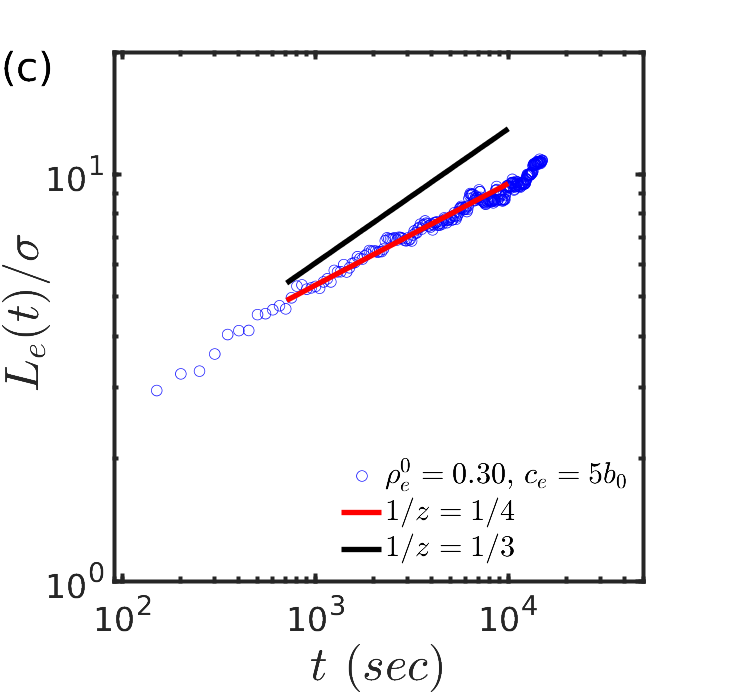}
    \includegraphics[width=0.32\textwidth]{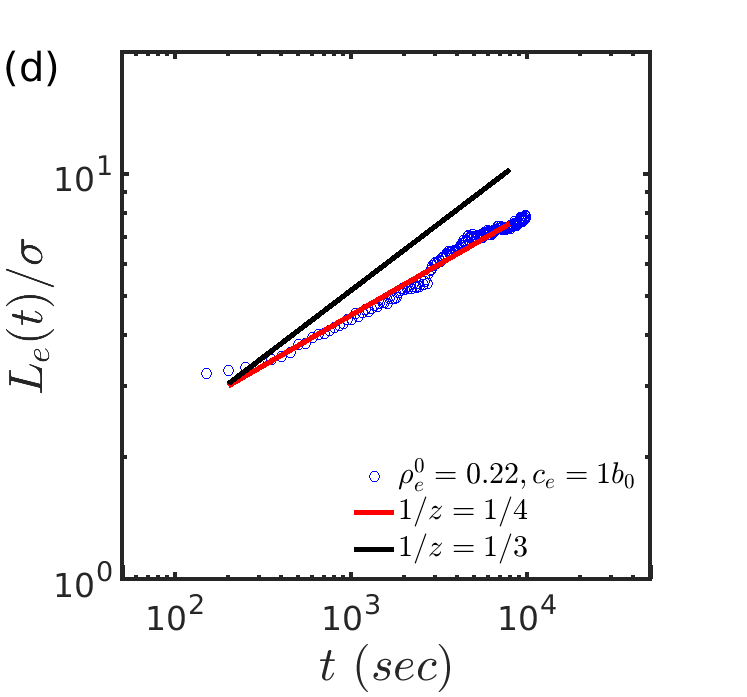}
    \includegraphics[width=0.32\textwidth]{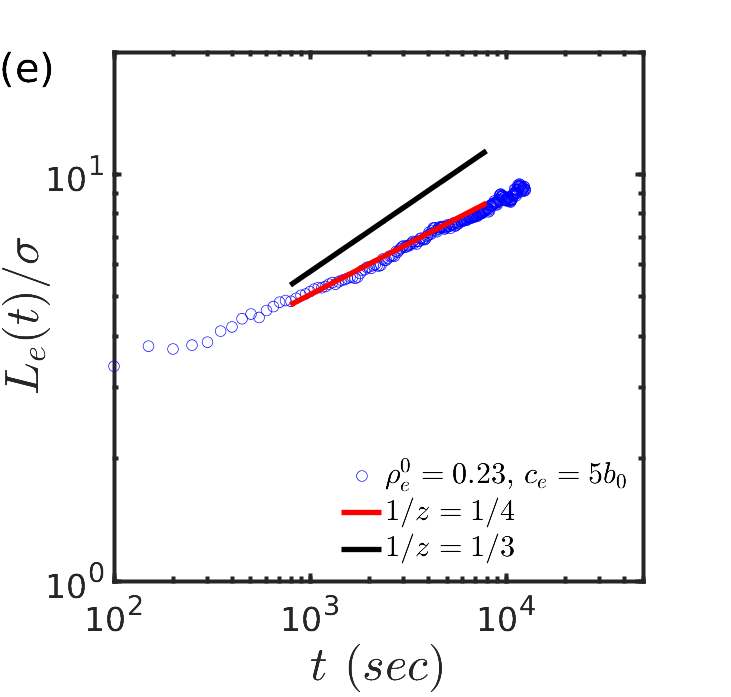}
    \includegraphics[width=0.32\textwidth]{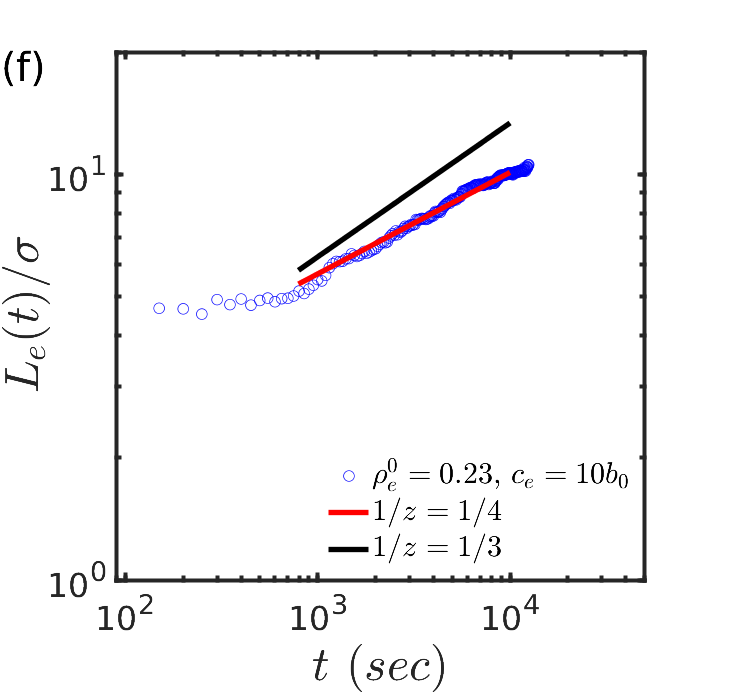}
    \caption{Upper panels (a)–(c): Length scale $L_e(t)$ at various area fractions of colloids corresponding to $\rho_e^0 = 0.1,~ 0.2,~ \text{and}~ 0.3$, respectively, and a fixed cell concentration of $c_e = 5b_0$. Lower panels (d)–(e): Length scale $L_e(t)$ at various concentrations of bacteria $c_e = 1b_0,~ 5b_0,~ \text{and}~ 10b_0$ for a fixed area fraction $\rho_e^0 \sim 0.2$. The red and black lines have slopes of $1/4$ and $1/3$, respectively.}
    \label{len_scale}
\end{figure}

To confirm the robustness of the coarsening exponent associated with the growth of colloidal clusters, we have determined the length scale $L_e(t)$ for various area-fractions of colloids, $\rho_e^0 = 0.1$, $0.2$ and $0.3$, at a cell concentration of $c_e = 5b_0$, as shown in Figs.~\ref{len_scale}(a)–(c). We have also determined $L_e(t)$ for different bacterial concentrations, $c_e = 1b_0$, $5b_0$, and $10b_0$, at a fixed area fraction $\rho_e^0 \sim 0.2$ of colloids, as shown in Figs.~\ref{len_scale}(d)–(e). The red line corresponds to a slope of $1/4$, while the black line has a slope of $1/3$. The length scale $L_e(t)$ was extracted from the spatial correlation function of the order parameter $\phi$, as described in the main text. These results clearly demonstrate that a growth exponent of $\sim1/4$ is robust across a range of colloidal area fractions and cell concentrations.

\subsection{Effective diffusion constant, cross over time-scale and length-scale}

Table~\ref{table1} shows the effective diffusion coefficient $D_{\rm eff}$, the crossover time $t_c$ and the length $r_c$ measured in our experiments for colloids of size $15~\mu m$ over a range of bacterial concentrations $c_e=1-10$. \\

{The crossover time $t_c$ represents the characteristic duration over which the motion of the passive particles remains correlated due to the collective activity of the bacteria. For times $t < t_c$, successive bacterial kicks are correlated, causing a particle motion to persist nearly in the same direction and exhibit superdiffusive motion. For $t > t_c$, this correlation is lost because the bacterial configuration has reorganized, and the particle experiences effectively uncorrelated kicks, leading to normal diffusion. Thus, $t_c$ can be interpreted as the persistence time of the bacteria-driven particle motion.}\\

The variation of $c_e$ by an order of magnitude leaves the characteristic time scale unchanged. The experimental length scale $r_c$, estimated as $(D_{\rm eff}~t_c)^{1/2}$, changes from $0.74$ to $1.29$ as $c_e$ varies from 1 to 10.\\

\begin{table}[h!]
    \centering
    \begin{tabular}{|l|l|l|l|}
    \hline
    $c_e$~($b_0$) & $D_{\rm eff}$~($\mu m^2/sec$) & $t_c$~($sec$) & $r_c$~$(\mu m$) \\
    \hline
    1 & 0.911 & 0.6 &  0.7393 \\
    \hline
    5 & 1.429 & 0.6 & 0.9260 \\
    \hline
    10 & 2.395 & 0.7 & 1.2948 \\
    \hline
    \end{tabular}
    \caption{The effective diffusion coefficient $D_{\rm eff}$, the crossover time $t_c$ and the crossover mean square length $r_c$ for colloids of size $15~\mu m$ at various concentrations of bacteria $c_e$.}
    \label{table1}
\end{table}

\newpage
\section{Theoretical modeling}

\subsection{Model B with spatio-temporal colored noise}

The conserved order parameter corresponding to the density field is defined as 
$\phi(\boldsymbol{r},t) = 2 \rho(\boldsymbol{r},t) -1 $. The evolution of $\phi(\boldsymbol{r},t)$ is governed by the continuity equation, 
\begin{align}
     &\quad \frac{\partial \phi(\boldsymbol{r},t)}{\partial t} = -\boldsymbol{\nabla} \cdot \boldsymbol{J} +\boldsymbol{\nabla} \cdot (\sqrt{2D(c)}\boldsymbol{\xi}) .
     \label{eq:4}
\end{align}
Recent experiments have shown that the effective diffusivity of passive colloids varies linearly with the mean concentration of swimmers in the suspension \cite{Libchaber00, Arratia16}, hence we set $D(c) \sim c$.  The first term in Eq.~\eqref{eq:4} is the current 
$\boldsymbol{J} (\boldsymbol{r},t) = -D(c)\boldsymbol{\nabla}\mu$, 
which is driven by the ``chemical potential'' $\mu$, and the second term denotes the spatio-temporal colored noise of the active bath. The chemical potential takes the form $ \mu = \delta \mathcal{F} / \delta \phi$, where the ``free energy functional'' is expressed in the standard Ginzburg-Landau form \cite{Bray94, Lubensky00} in dimensionless units as
\begin{equation*}
    \mathcal{F}[\phi] = {\int} \left[ -\frac{1}{2} \phi(\boldsymbol{r},t)^2  + \frac{1}{4} \phi(\boldsymbol{r},t)^4 + \frac{1}{2} (\nabla\phi(\boldsymbol{r},t))^2 \right] d \boldsymbol{r}.
\end{equation*}
where, the first two terms constitute the local bulk free-energy density and generate a symmetric double-well potential, which drives phase separation and determines the equilibrium coexistence densities of the two homogeneous phases in model B. The square-gradient term penalizes spatial variations of the order parameter, thereby regularizing the interface, producing a finite interfacial width, and giving rise to a nonzero surface tension \cite{Bray94, Lubensky00}.

\subsection{Numerical details}

We numerically integrate equation of Model B (MB) with colored noise 
\begin{align}
     \quad \frac{\partial \phi(\boldsymbol{r},t)}{\partial t} = \nabla^2\left(-\phi + \phi^3 - \nabla^2 \phi\right) +\boldsymbol{\nabla} \cdot \boldsymbol{\xi},
     \label{eq:modelb}
\end{align}
where the simplest spatio-temporally correlated colored noise $\xi (\boldsymbol{r},t)$ can be obtained by solving the linear second order reaction-diffusion equation \cite{garcia1992generation,sagues2007spatiotemporal}:
\begin{equation}
\frac{\partial}{\partial t} \xi_i(\boldsymbol{r},t) = -\frac{1}{\tau}(1-\lambda^2\nabla^2) \xi_i(\boldsymbol{r},t) + \frac{1}{\tau}\eta_i(\boldsymbol{r},t) .
\label{noise}
\end{equation}
{ Dimensionless space and time in Eqs.~(\ref{eq:modelb})-(\ref{noise}) are measured in units of the interfacial width $\xi_b$ and the diffusion time $\xi_b^2/D$, respectively \cite{Puri09}. Assuming a single length scale, these are proportional to $r_c$ and $t_c$ in Table~\ref{table1}. However, the corresponding proportionality constants cannot be determined unambiguously, since the coarse-grained description does not retain all the microscopic details of the underlying systems.} The subscript $i$ takes values 1,2 for the two components of vector ${\bf \xi}$. The persistence length $\lambda$ and the time scale $\tau$ of the noise arise from the Laplacian term, which couples the values of the field $\xi({\bf r}, t)$ at different points. In Eq.~(\ref{noise}), $\eta(\boldsymbol{r},t)$ is a Gaussian white noise with zero average and correlation
\begin{equation}
\langle \eta_i(\boldsymbol{r},t) \eta_j(\boldsymbol{r'},t') \rangle = 2\epsilon \delta_{ij}\delta(\boldsymbol{r}-\boldsymbol{r'}) \delta (t-t') , 
\label{noise2}
\end{equation}
where the strength $\epsilon = 5$ in our simulations. We also checked the results for larger $\epsilon = 25, 50$ and this only affects prefactors.  \\

The Eqs.~(\ref{eq:modelb})  and ~(\ref{noise}) are numerically integrated using the Euler scheme with $\Delta x = 1.0$ and $\Delta t = (0.001-0.005)$. The system is a two dimensional box of size $K \times K$ with periodic boundary condition in both directions. 
The size of the system $K$ is varied from $256-2048$, and the total simulation time $t_s = [5\times 10^4-2\times 10^7]$ with the actual time being $t_s \times \Delta t$. One simulation time consists of a single update of the equations for $\phi$ Eq. ~(7) (in the main manuscript) and $\xi$ in Eq.~(\ref{noise}) for all lattice points. The magnitudes of $\lambda$ and $\tau$ in our simulations are chosen in light of the experimental input. The persistence time of colloidal beads shows minimal change, whereas their persistence length experiences small but noticeable variations with bacterial concentration. To mimic the effect of variation in bacterial concentration, the correlation time $\tau$ is varied between $0.1$ and $2.0$, and $\lambda$ is varied from $0$ to $9$. { Notice that this range overestimates the variation seen in our experiments in Table~\ref{table1}. This was done to confirm the robustness of our numerical results.} The majority of the results shown in the main manuscript are for $\tau = 1.0$ and the system size $K = 1024$, and some results for $K = 2048$ are shown in the supplementary material. \\

We start from a disordered homogeneous state at $t=0$ with small-amplitude fluctuations: $\phi (\boldsymbol{r},0) = \phi_0 + \delta \phi (\boldsymbol{r},0)$. The corresponding mean density is $\rho_0 = (1+\phi_0)/2$. The coupled system of partial differential equations for $\phi$ and $\xi$ are evolved to obtain configurations of the order parameter. The ordering kinetics is characterized by the evolution of morphologies, correlation functions, and structure factors as defined in Eqs.~(1) and (2) in the main manuscript. The mean density of the particles is kept to low values $\rho_0 = 0.05, 0.1, 0.15$ with the corresponding values of $\phi_0 = -0.9, -0.8, -0.7$, respectively. All statistical quantities reported here are obtained as an average over $50-100$ independent runs.

\subsection{Results of Model B (MB)}

\begin{figure}[h!]
    \begin{tabular}{cc}

     \includegraphics[width=0.45\linewidth]{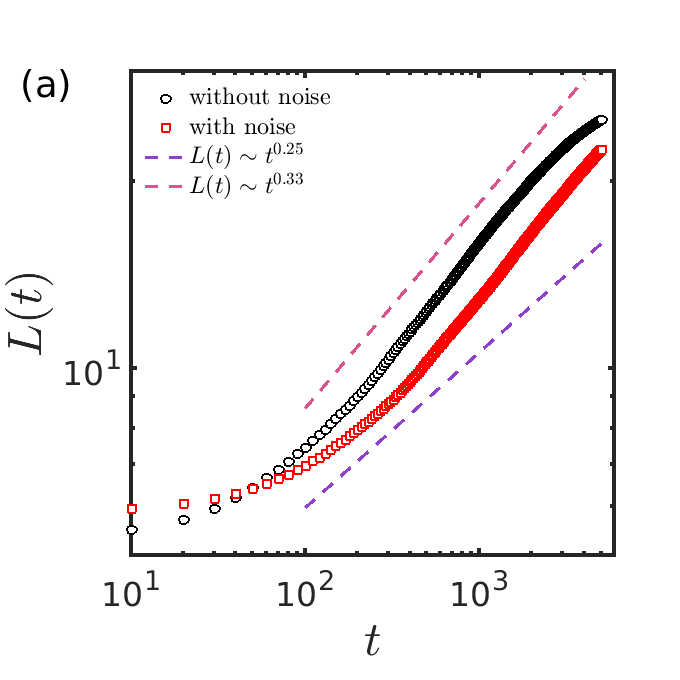} &
     \hspace{-1.5cm}

     \raisebox{0.15cm}{%
     \includegraphics[width=0.57\linewidth]{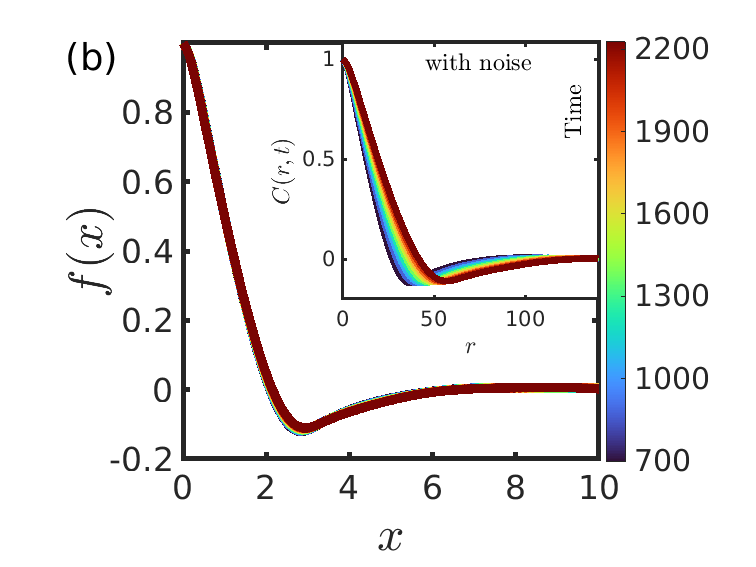}} \\

     \includegraphics[width=0.57\linewidth]{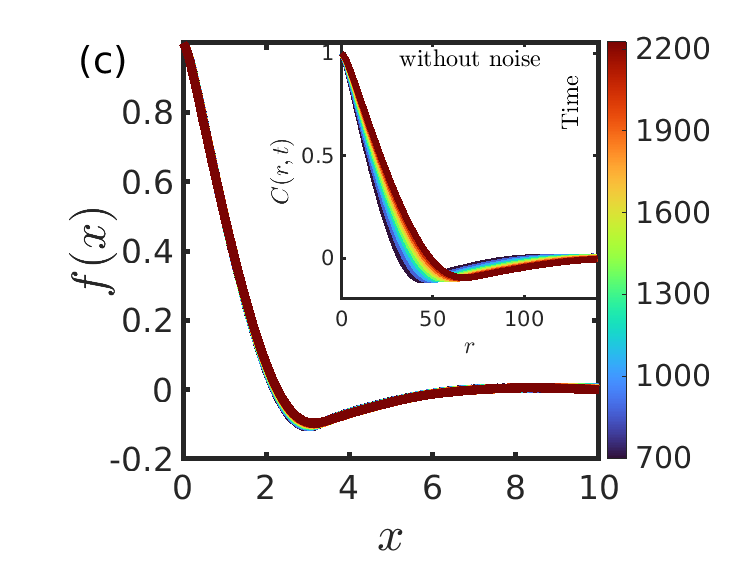} &
     \hspace{-1.5cm}

     \raisebox{-0.15cm}{%
     \includegraphics[width=0.45\linewidth]{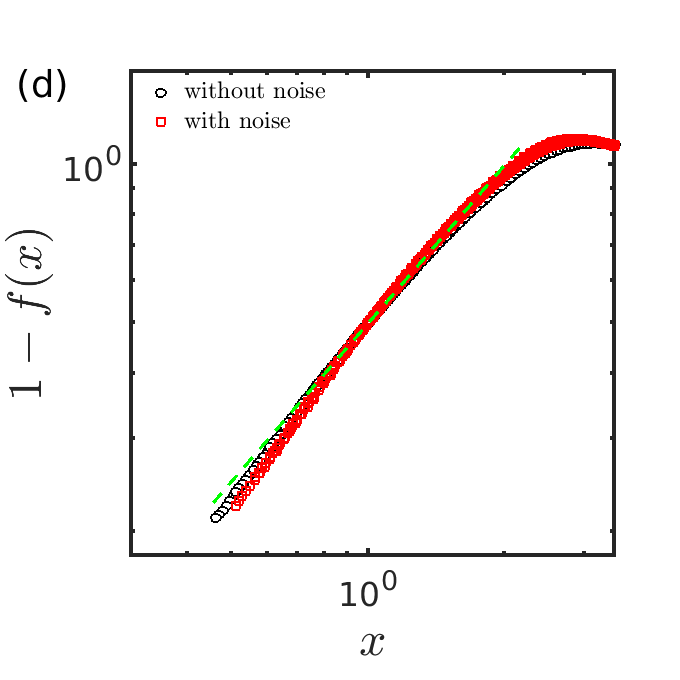}} \\

    \end{tabular}
    \caption{Ordering kinetics in model B at density $\rho_0 = 0.15$ with and without Gaussian white noise. Panel (a) shows the time evolution of characteristic length scale $L(t)$. For both cases, the growth of $L(t)$ is consistent with $t^{1/3}$. The panels (b-c) display the dynamic scaling of correlation function, $C(r,t)$, with noise and without noise, respectively. The main panels in (b) and (c) shows scaling collapse when the distance is scaled as $rt^{-1/3}$. Panel (d) shows the plot of the reduced scaling function ($1-f(x)$) showing the cusp singularity of the correlation function. Line of slope $1$ is shown as a dashed line.}
    \label{fig:mbld}
\end{figure}

Figure~\ref{fig:mbld} shows the coarsening dynamics of Model B at low density ($\phi_0 = -0.70$ or $\rho_0=0.15$), both in the absence and in the presence of Gaussian white noise. For the latter case, the noise amplitude is fixed at $0.10$. In both cases, the characteristic domain size exhibits the classical Lifshitz--Slyozov growth law, $L(t) \sim t^{1/3}$, as demonstrated in Fig.~\ref{fig:mbld}(a). The emergence of a well-defined scaling regime for both cases (with and without noise) is further confirmed by the collapse of the two-point correlation function at different times upon appropriate rescaling ($x=\frac{r}{L(t)}$), as shown in Fig.~\ref{fig:mbld}(b) \& (c). 
To investigate interfacial properties, we examine the reduced scaling function $1 - f(x)$ as a function of the scaled distance $x$ [Fig.~\ref{fig:mbld}(d)]. The linear behavior observed at small $x$ is consistent with sharp interfaces and indicates the absence of interface roughening. \\

\subsection{Microphase separation}
\begin{figure}[h!]
    \centering
    \includegraphics[width=0.90\linewidth]{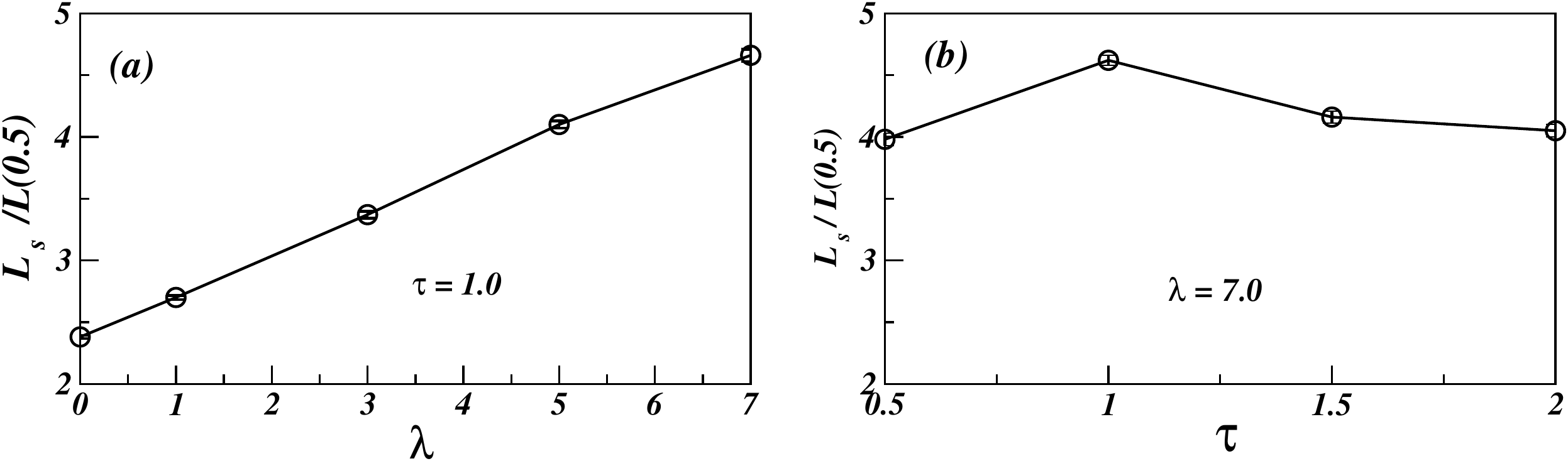}
    \caption{(a) Normalized saturation length $L_s/L(0.5)$ vs. $\lambda$ for $\tau =1$. (b) $L_s/L(0.5)$ vs. $\tau$ for $\lambda = 7$. The other parameters are the same as for Fig. 3(b-c) in the main manuscript.}
    \label{sfig14}
\end{figure}

To confirm microphase separation in simulation, we show the variation of (a) the normalized saturation length $L_s/L(0.5)$ vs. $\lambda$ for fixed $\tau = 1$, and (b) $L_s/L(0.5)$ vs. $\tau$ for $\lambda = 7$ in Fig.~\ref{sfig14}. The finite values of $L_s$ confirm microphase separation for a wide range of parameters. On dimensional grounds, we expect $L_s \sim \lambda$ and $L_s$ to be independent of $\tau$.
 
\subsection{Finite size analysis}

\begin{figure}[h]
   \centering
\includegraphics[width=0.42\textwidth]{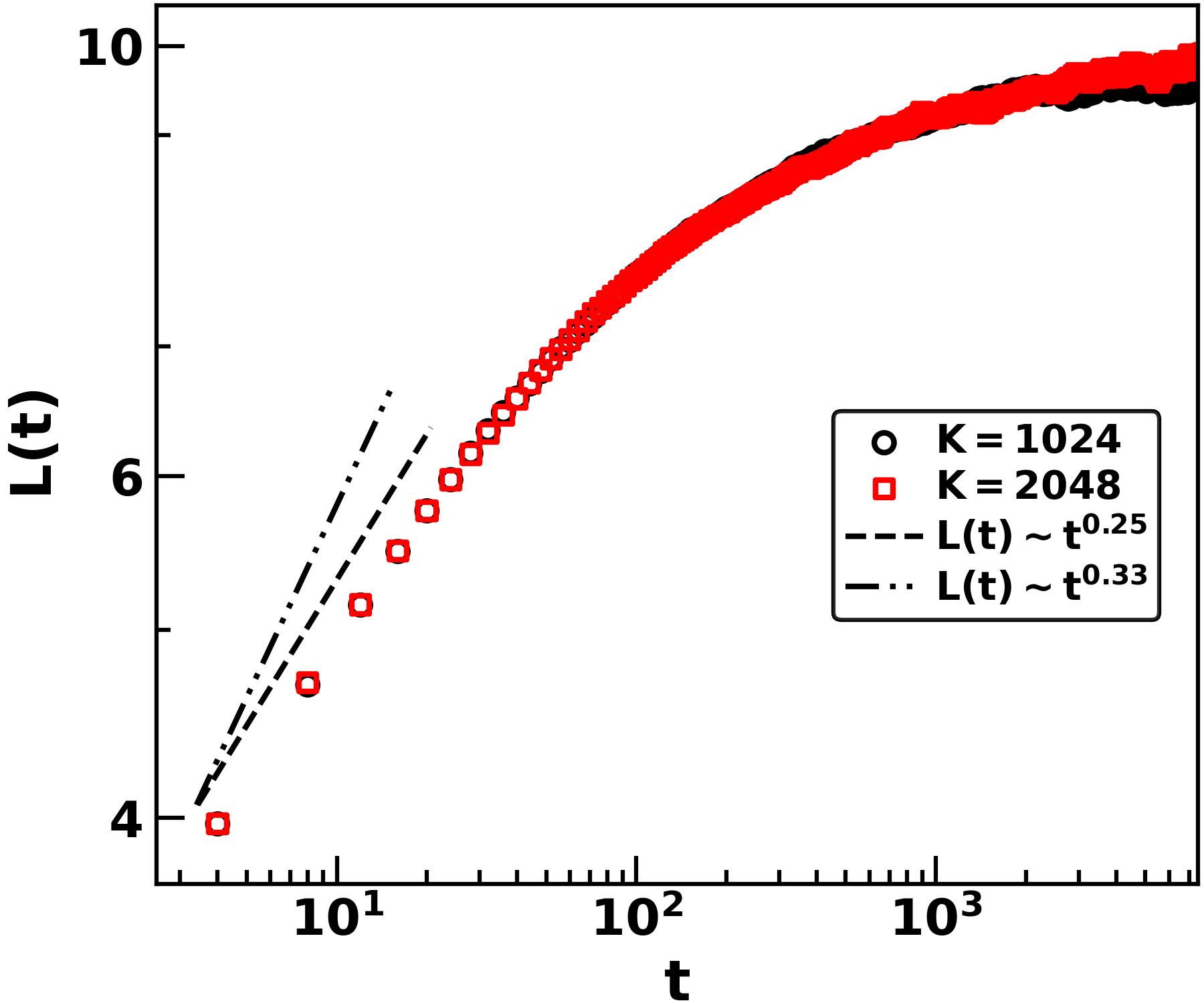}

\caption{The plot presents the finite size analysis of $L(t)$ vs. $t$ for two system sizes $K = 1024$ and $2048$ keeping $\lambda =7, \tau=1$, and $\rho_0 = 0.15$. For both system sizes at late times, $L(t)$ saturates nearly to the same value. The dashed and dash-dotted lines have slopes of $1/4$ and $1/3$, respectively.}
\label{fig_fsize}
\end{figure}

Our simulations reveal that domain growth is arrested on a longer time scale, leading to saturation of the domain size. However, such a saturation can also arise due to finite size effects. In Fig. \ref{fig_fsize} we show $L(t)$ vs. $t$ for two different system sizes $(K = 1024,~2048)$. At early times the $L(t) \sim t^{1/4}$ and shows a clear saturation at late times, with saturation length $<< K$. Hence, such saturation is not due to the finite size effect. 

\subsection{Dynamic scaling of the correlation function in simulations}
\begin{figure}[h!]
    \centering
    \includegraphics[width=0.35\linewidth]{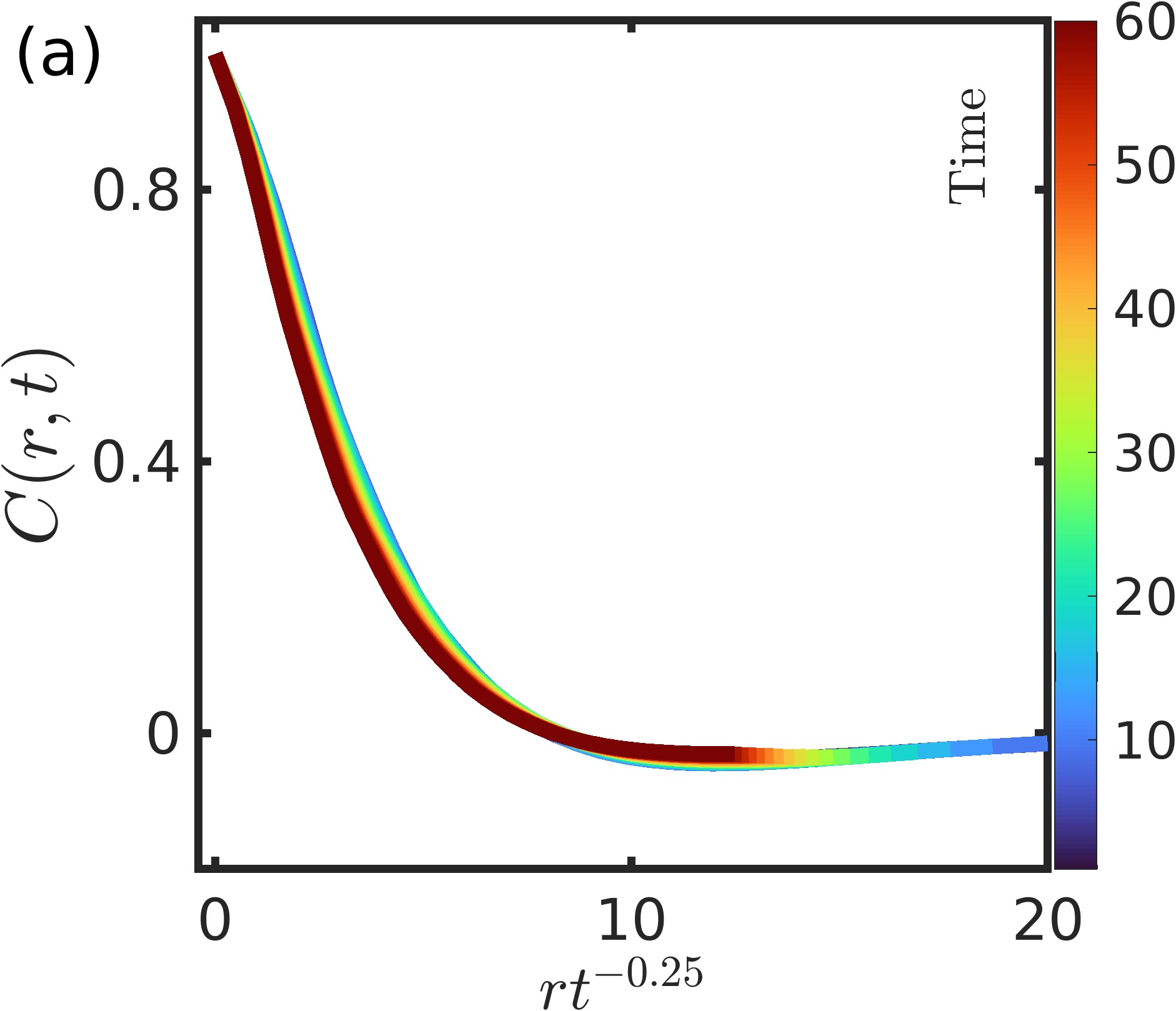}
    \includegraphics[width=0.35\linewidth]{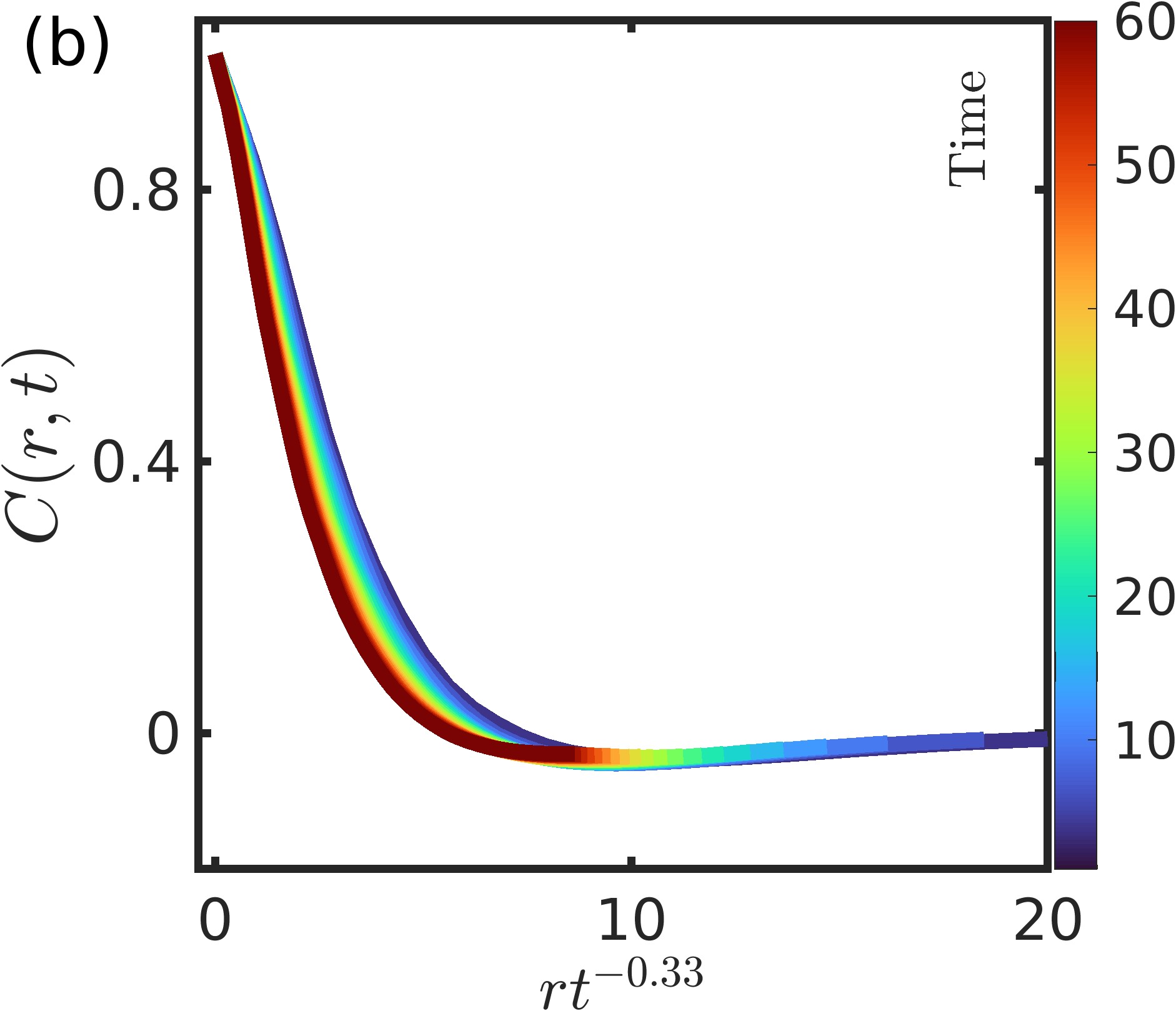}
\caption{Dynamic scaling and slow growth of the correlation function $C(r,t)$. The function $C(r,t)$ in panels (a) and (b) are scaled by $t^{1/4}$ and $t^{1/3}$, respectively. The color bar represents the time progression.  Here, $\lambda = 7$ and $\tau = 1$. The mean density of the system is $\rho_0 = 0.15$. The data are obtained from a simulation box of size $K = 1024$.}
\label{sfig4}
\end{figure}

Fig.~\ref{sfig4} presents two different scaling plots of the correlation function $C(r,t)$. Data collapse is clearly better when $C(r,t)$ is scaled using $t^{1/4}$, instead of $t^{1/3}$.

\subsection{Robustness of slow growth and rough interfaces }

\begin{figure}[hbt]
\centering
\includegraphics[width=0.75\linewidth]{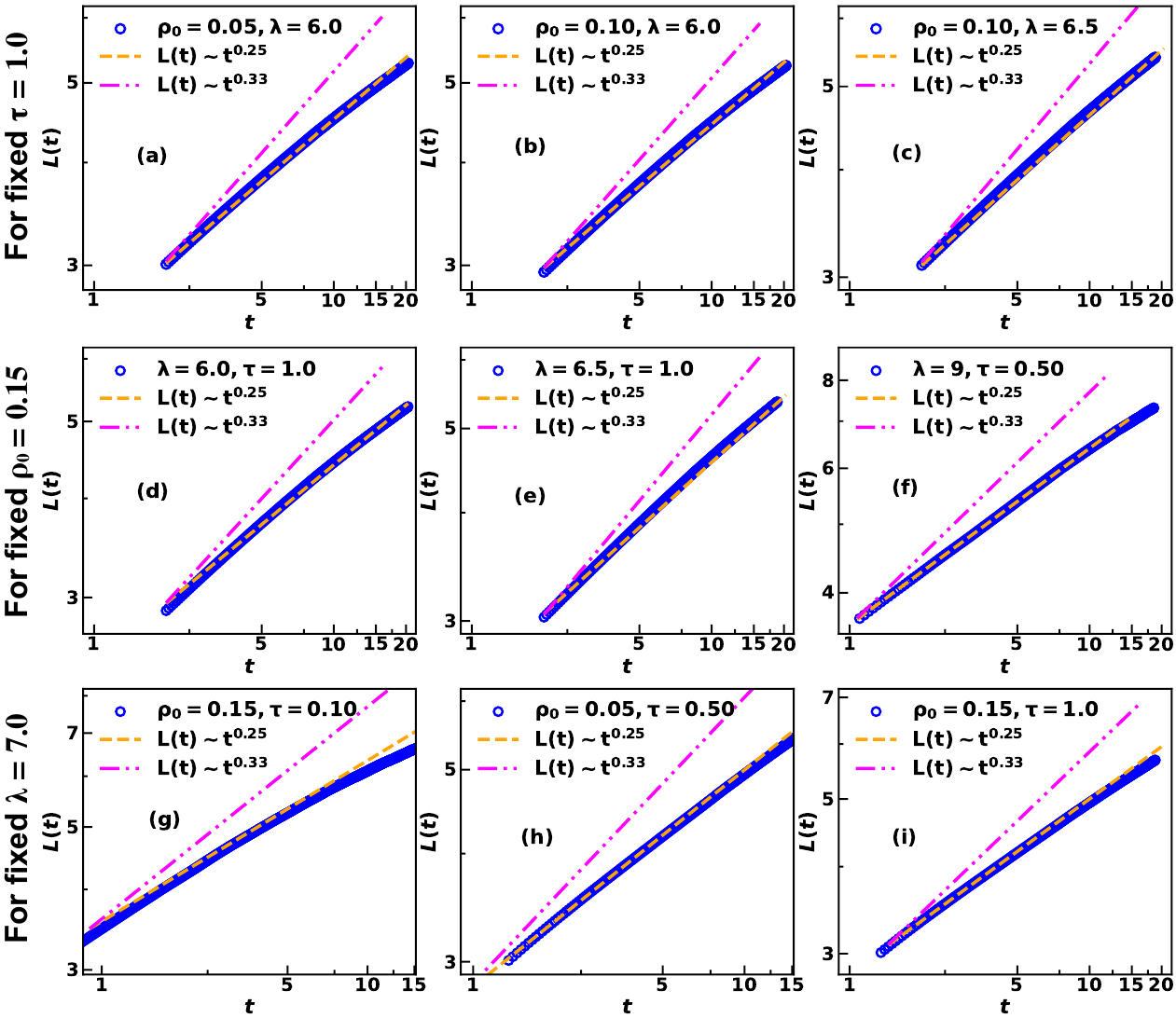}
\caption{Time evolution of $L(t)$ for different choices of control parameters. Panels (a-c) show the domain growth $L(t)$ \emph{vs.} $t$ at a fixed $\tau = 1.0$ for three different combinations of $(\rho_0,\lambda)$: (a) $(0.05,6.0)$, (b) $(0.10,6.0)$, (c) $(0.10,6.5)$; Panels (d-f) show the $L(t)$ \emph{vs.} $t$ plot at a fixed $\rho_0=0.15$ for three different combinations of $(\lambda,\tau)$: (d) $(6.0,1.0)$, (e) $(6.5,1.0)$, (f) $(9.0,0.50)$; Panels (g-i) show the $L(t)$ \emph{vs.} $t$ plot at a fixed $\lambda=7.0$ for three different combinations of $(\rho_0,\tau)$: (g) $(0.15,0.10)$, (h) $(0.05,0.50)$, (i) $(0.15,1.0)$. All the plots shown are for a system size $K=2048$.}
\label{sfig7}
\end{figure}

In this section, we discuss the robustness of the growth law found in our simulations. The Figs.~\ref{sfig7}(a-c) display the growth of $L(t)$ at $\rho_0=\textnormal{0.05, 0.1}$ and $\lambda= (6.0, 6.5)$ with fixed $\tau  = 1.0$, while the panels in (d-f) present results for $\lambda=(6.0-9.0)$ with fixed $\rho_0 = 0.15$ and $\tau = 1.0$ and $0.5$. The Figs.~\ref{sfig7}(g-i) presents $L(t)$ for fixed $\lambda = 7$ and varying $\tau$ and $\rho_0$. It is clear from these results that growth exponent remains close to $1/4$ when the parameters $\rho_0$, $\lambda$ and $\tau$ are varied.\\

\begin{figure}[hbt]
    \centering
\includegraphics[width=0.7\linewidth]{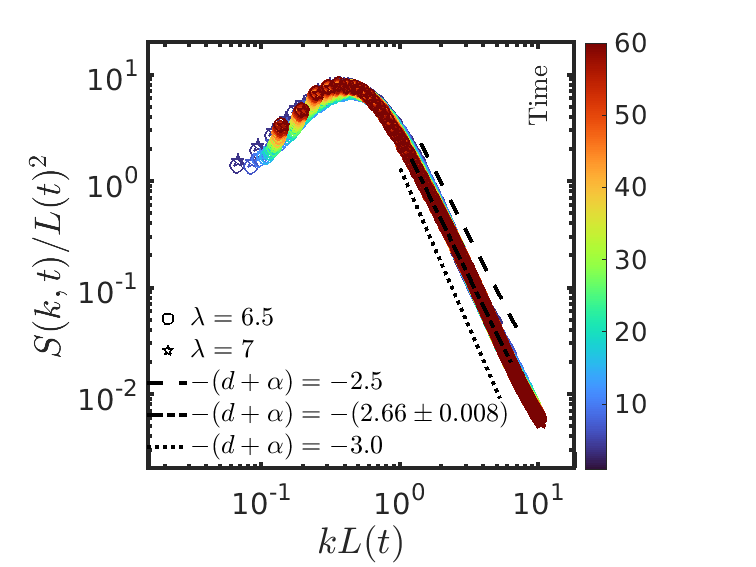}
\caption{The scaled structure factor $L^{-2}(t)S(k)$ is plotted against $k L(t)$ for  Model B with colored noise for $\lambda = 6.5$ and $7.0$, and $(\rho_0, \tau)=(0.15,1.0)$. The system size is $K=1024$.}
\label{sf_comparison}
\end{figure}

We also demonstrate that the interfaces remain rough over a range of parameters investigated in our simulations. The variation of growth and cusp exponents for a broad range of parameters are given in Table \ref{tab:sample}. Figure~\ref{sf_comparison} shows the scaled structure factor $L^{-2}(t)S(k)$ for $\lambda=\textnormal{6.5 and 7.0}$. The strong non-Porod scaling of the structure factor shows that rough interfaces are characteristic features of phase separation in the presence of colored noise. These results are supported by the visual impression of the domains presented in Fig.~\ref{sfig5}.   

\begin{table}[h]
\centering
\begin{tabular}{|c|c|c|c|c|}
\hline
$\rho_o$ &$\lambda$ & $\tau$ & Growth exponent & Cusp exponent \\

\hline
0.05 & 6 & 1 & $0.24756 \pm 0.00047$ & $0.6619 \pm 0.0092$\\
0.05 & 7 & 1 & $0.238 \pm 0.00021$ & $0.659\pm 0.013$\\
0.05 & 9 & 0.50 & $0.2522 \pm 0.00013$ & $0.652\pm 0.011$\\
\hline
0.10 & 6 & 1 & $0.258 \pm 0.0004$ & $0.656 \pm 0.0012$\\
0.10 & 7 & 1 & $0.249 \pm 0.00035$ & $0.673 \pm 0.0061$\\
0.10 & 9 & 0.50 & $0.247 \pm 0.00042$ & $0.666 \pm 0.009$\\
\hline
0.15 & 6 & 1 & $0.265 \pm 0.0004$ & $0.669 \pm 0.0083$\\
0.15 & 7 & 2 & $0.251 \pm 0.0003$ & $0.659 \pm  0.0012$\\
0.15 & 7 & 0.10 & $0.261 \pm 0.0005$ & $0.665 \pm 0.0075$\\
0.15 & 9 & 0.50 & $0.256 \pm 0.0003$ & $0.667 \pm 0.009$\\

\hline
\end{tabular}
\caption{Growth and cusp exponents for a range of parameters $(\rho_0, \lambda, \tau)$. The data are obtained for a system size $K = 1024$.}
\label{tab:sample}
\end{table}

\section{Comparison of evolution of order parameter in experiments and simulations}

\begin{figure}[hbt]
\centering
\includegraphics[width=1\linewidth]{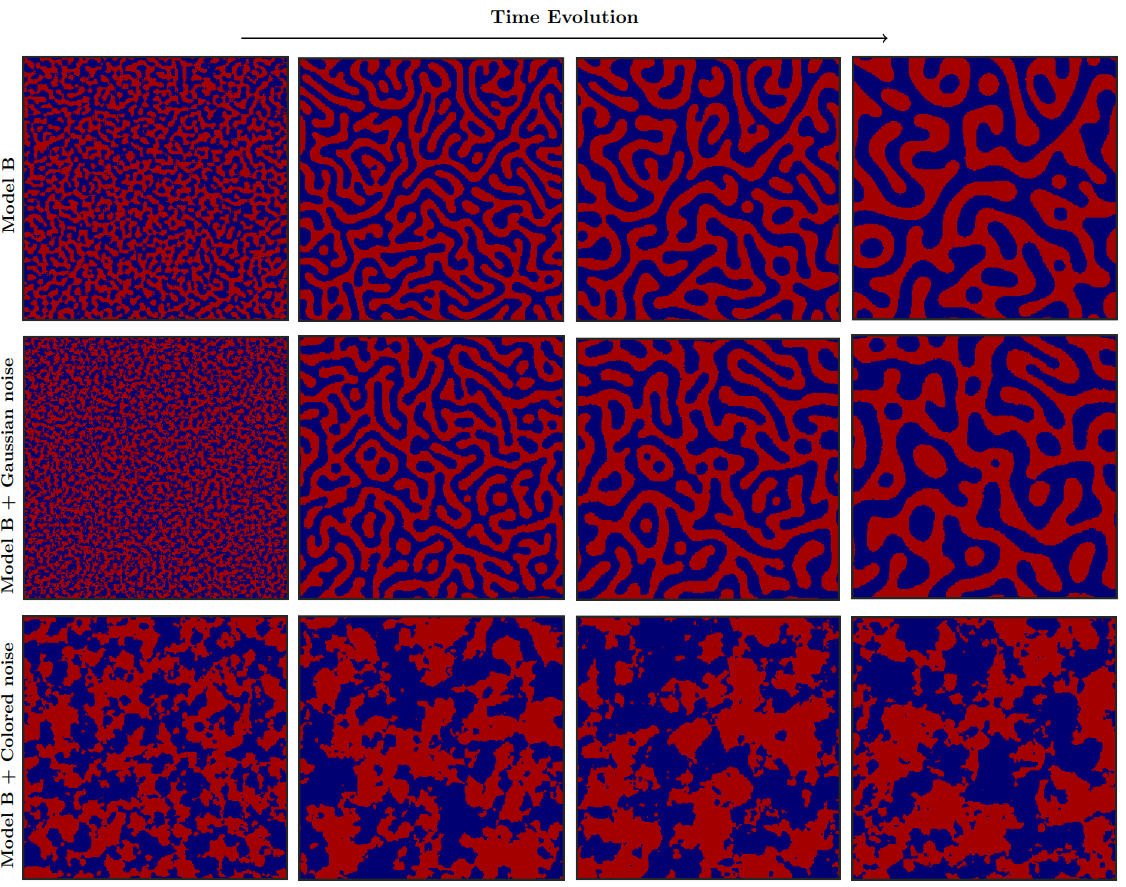}
\caption{Comparison of the evolution of the hardened order parameter field $\phi(\mathbf{r},t)$ in simulations. The panels show the time evolution of $\phi(\mathbf{r},t)$ for Model B at the critical composition, $\phi_0 = 0$ ($\rho_0=0.5$). Panel (a) corresponds to the noiseless case, panel (b) is with the inclusion of Gaussian white noise, and panel (c) is with the inclusion of colored noise with parameters $\tau = 1$ and $\lambda = 7$. The columns represent snapshots at times $t = 2$, $150$, $400$, and $1000$. All snapshots are shown for a system size of $K = 256$. Blue and red regions correspond to $\phi= -1$ and $\phi= +1$, respectively.}
\label{sfig6}
\end{figure}

We provide a comparison of the order parameter in experiments and simulations in this section. In Fig.~\ref{sfig6}, we show instantaneous snapshots of phase-separating domains for Model B with a critical composition ($\phi_0 = 0$ or $\rho_0=0.5$) in three different cases: (i) MB without thermal noise, (ii) MB with Gaussian white noise, and (iii) MB with colored noise $(\tau = 1, \lambda = 7)$. In Fig.~\ref{sfig6}, from left to right, we show snapshots for different times: $2, 150, 400, 1000$ and for $K = 256$. The domain morphology is very different for MB with colored noise in comparison to cases (i) and (ii). In particular, the domain walls are rough, suggesting a fractal nature. This is reflected in the non-Porod behavior of the correlation functions and structure factors, shown in the main manuscript.\\

\begin{figure}[h]
\centering
\includegraphics[width=1\linewidth]{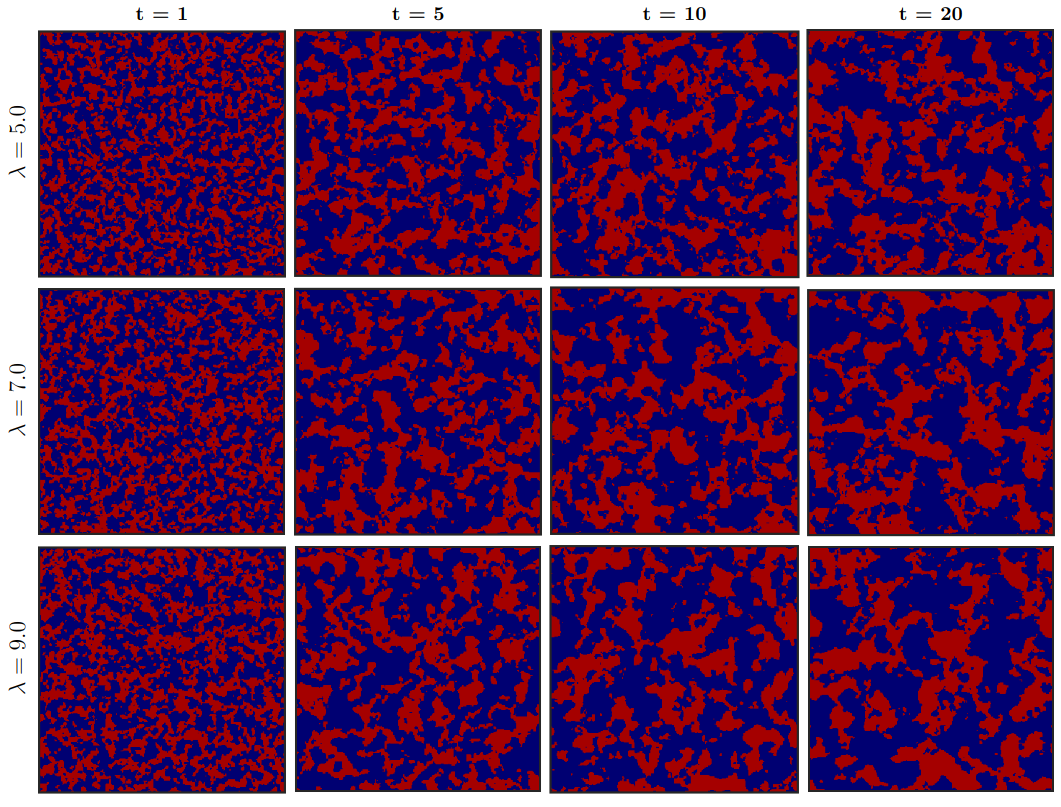}
\caption{Phase diagram in the plane of $\lambda$ and time in the simulations. The panels display the time evolution of the hardened order parameter field $\phi(\boldsymbol{r},t)$. The correlation length $\lambda$ is varied over the range $\lambda = 5-9$. The mean density of the system is $\rho_0 = 0.15$. Each snapshots is for $\tau = 1.0$ and $K = 256$. The blue and red colors indicate $\phi=-1$ and $\phi=+1$, respectively.}
\label{sfig5}
\end{figure}

In Fig.~\ref{sfig5}, we show the evolution of hardened local order parameter $\phi(\boldsymbol{r},t)$. The red and blue denote regions of high and low density, respectively. From left to right in the figure, the snapshots are depicted at different times for a mean density $\rho_0 = 0.15$ $(\phi_0 = -0.7)$. From top to bottom, we show snapshots for the system with  colored noise for different $\lambda = (5, 7, 9)$ at times: $1, 5, 10, 20$ and $K = 256$.   \\

\begin{figure}[h]
    \centering
    \includegraphics[width=1\linewidth]{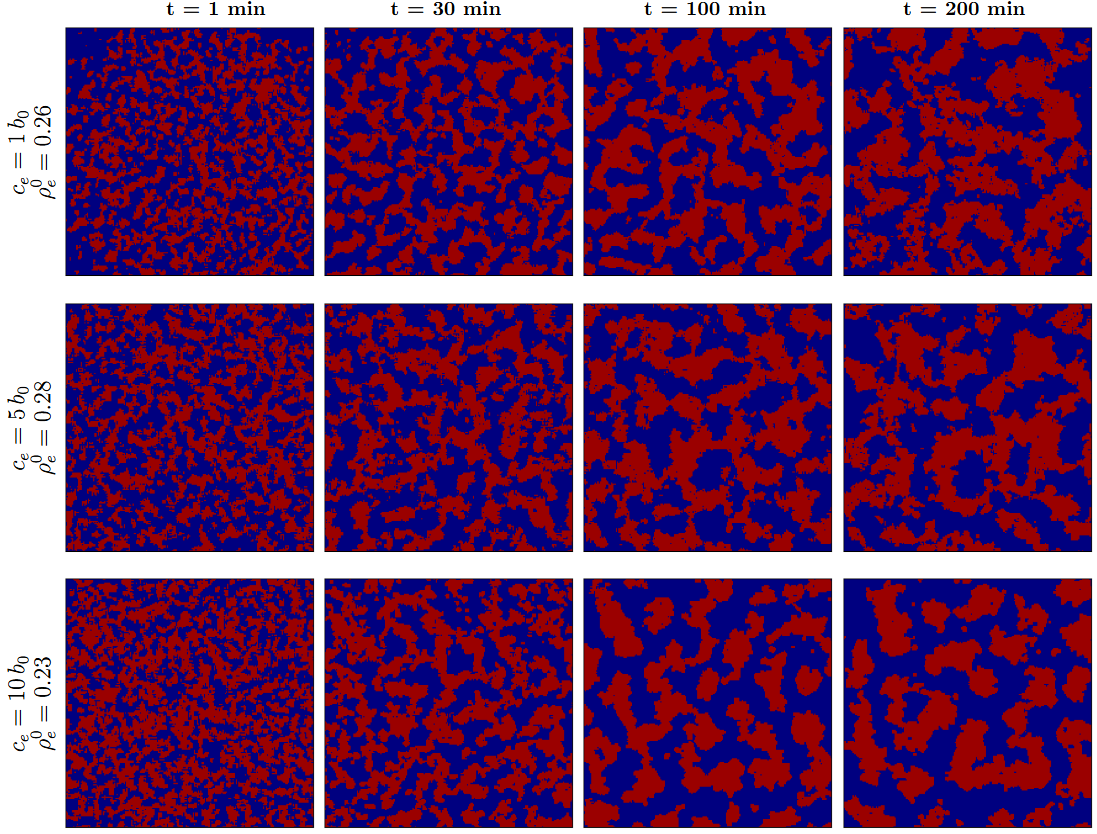}
    \caption{The panels display the time evolution of the order parameter $\phi_e(\boldsymbol{r},t)$ in experiments in the plane of bacterial concentration $c_e$ and time. From left to right, the panels depict the evolution of local order parameter at $t=1$min, $30$min, $100$min and $200$min. The concentration of microswimmers is varied from $c_e=1-10$. The mean density of colloids in the top, middle and bottom panels is  $\rho_e^0 = 0.26$, $0.28$ and $0.23$, respectively. The blue and red colors indicate $\phi=-1$ and $\phi=+1$, respectively.}
    \label{fig:placeholder}
\end{figure}

\clearpage
\newpage

\section{Movies of the coarsening of colloidal domains in experiments and simulations}

The supplementary information contains videos of the coarse-grained hardened order parameter field $\phi(\boldsymbol{r},t)$, obtained from both experiments and simulations.  The order parameter, $\phi$, is color-coded. The information about the video files is given below. \\

SM1: Formation of colloidal domains in experiments over a duration of $210$ min. The size of the colloids is $\sigma=15\mu m$, the area fraction of the colloids is $\rho_e^0=0.28$ and the concentration of the swimmers is $c_e=5b_0$. The playback speed has been accelerated to present the slow process of colloidal ordering. \\

SM2: Colloidal domains formation in experiments over a duration of $210$ min. The size of the colloids is $\sigma=10\mu m$, the area fraction of the colloids is $\rho_e^0=0.27$ and the concentration of bacteria is $c_e=5b_0$. The playback speed has been accelerated to present the slow process of colloidal ordering.\\

SM3: Phase separation kinetics of colloidal domains in experiments. The video captures the evolution of hardened order parameter field $\phi_e(\mathbf{r},t)$ over a time scale of $210$ minutes, beginning from the homogeneous phase at the start of the experiment. Red and blue indicate colloid-rich ($\phi=+1$) and colloid-poor ($\phi=-1$) phases, respectively. The bacterial concentration is $c_e = 5b_0$, and the colloidal area fraction is $\rho_e^0 = 0.28$. \\

SM4: Phase separation kinetics of domains in simulations with spatio-temporally correlated noise. The video shows the evolution of the system from the initial homogeneous phase at $t = 0$ to the phase-separated state at $t = 30$. The hardened order parameter field $\phi(\mathbf{r},t)$ is color-coded. Blue and red regions correspond to $\phi(\mathbf{r}) = -1$ and $\phi(\mathbf{r}) = +1$, respectively. The parameters used in the simulation are: average density $\rho_0 = 0.15$, system size $K = 512$, noise correlation length $\lambda = 7$, and correlation time $\tau = 1$. \\

SM5: Late time video of domains in simulations with spatiotemporal noise ($t=30$ to $60$). Other parameters are the same as in SM4. \\

SM6: Phase separation kinetics of domains in Model B with Gaussian white noise. The video shows the evolution of the system from $t = 40$ to $t = 60$ for a critical composition $\phi_0 = 0$ and system size $K = 512$. Blue and red regions correspond to $\phi = -1$ and $\phi= +1$, respectively.\\



%


\end{document}